\documentclass[conference,10pt]{IEEEtran}
\IEEEoverridecommandlockouts
\usepackage{cite}
\usepackage{url}
\usepackage{amsmath,amssymb,amsfonts}
\usepackage{algorithmic}
\usepackage[pdftex]{graphicx}
\usepackage{textcomp}
\usepackage{multirow}
\usepackage{xcolor}
\usepackage{amsmath,amssymb,amsfonts}
\usepackage{algorithmic}
\usepackage{graphicx}
\usepackage{textcomp}
\usepackage{subcaption}
\usepackage{array, makecell}
\usepackage{pifont}
\usepackage{adjustbox}
\usepackage{listings}
\usepackage{colortbl}
\usepackage{array, makecell}
\usepackage{enumitem}
\usepackage{hyperref}

\setcellgapes{2pt}

\def\BibTeX{{\rm B\kern-.05em{\sc i\kern-.025em b}\kern-.08em
    T\kern-.1667em\lower.7ex\hbox{E}\kern-.125emX}}
\begin{document}
\sloppy

\title{Performance Evaluation of Advanced Features in CUDA Unified Memory} 

\author{\IEEEauthorblockN{Steven W. D. Chien}
	\IEEEauthorblockA{\textit{KTH Royal Institute of Technology} \\
		Stockholm, Sweden \\
		wdchien@kth.se}
	\and
	\IEEEauthorblockN{Ivy B. Peng}
	\IEEEauthorblockA{\textit{Lawrence Livermore National Laboratory} \\
	Livermore, USA \\
	peng8@llnl.gov}
	\and
	\IEEEauthorblockN{Stefano Markidis}
	\IEEEauthorblockA{\textit{KTH Royal Institute of Technology} \\
	Stockholm, Sweden \\
	markidis@kth.se}
}

\maketitle

\begin{abstract}
	CUDA Unified Memory improves the GPU programmability and also enables GPU memory oversubscription. Recently, two advanced memory features, \emph{memory advises} and asynchronous \emph{prefetch}, have been introduced. In this work, we evaluate the new features on two platforms that feature different CPUs, GPUs, and interconnects. We derive a benchmark suite for the experiments and stress the memory system to evaluate both in-memory and oversubscription performance. 
	
	The results show that \emph{memory advises}  on the Intel-Volta/Pascal-PCIe platform bring negligible improvement for in-memory executions. However, when GPU memory is oversubscribed by about $50\%$, using \emph{memory advises} results in up to $25\%$  performance improvement compared to the basic CUDA Unified Memory. In contrast, the Power9-Volta-NVLink platform can substantially benefit from \emph{memory advises}, achieving up to 34\% performance gain for in-memory executions. However, when GPU memory is oversubscribed on this platform, using \emph{memory advises} increases GPU page faults and results in considerable performance loss. The CUDA \emph{prefetch} also shows different performance impact on the two platforms. It improves performance by up to $50\%$ on the Intel-Volta/Pascal-PCI-E platform but brings little benefit to the Power9-Volta-NVLink platform.
\end{abstract}


\section{Introduction} \label{sec:introduction}
Recently, leadership supercomputers are becoming increasingly heterogeneous. For instance, the two fastest supercomputers in the world~\cite{top500}, Summit and Sierra, are both equipped with Nvidia V100 GPUs~\cite{8425458,jia2018dissecting} for accelerating workloads. One major challenge in programming applications on these heterogeneous systems arises from the physically separate memories on the host (CPU) and the device (GPU). Kernel execution on GPU can only access data stored on the device memory. Thus, programmers either need to explicitly manage data using the memory management API in CUDA or relying on programming systems, such as OpenMP~4.5~\cite{karlin2016early} and RAJA~\cite{hornung2014raja}, for generating portable programs. Today, a GPU can have up to 16 GB memory on top supercomputers while the system memory on the host can reach 256 GB. Leveraging the large CPU memory as a memory extension to the relatively small GPU memory becomes a promising and yet challenging task for enabling large-scale HPC applications.

CUDA Unified Memory (UM) addresses the challenges as mentioned above by providing a single and consistent logical view of the host and device memories on a system. UM uses the virtual memory abstraction to hide the heterogeneity in GPU and CPU memories. Therefore, pages in the virtual address space in an application process may be mapped to physical pages either on CPU or GPU memory. Based on UM, CUDA runtime can leverage page faults, which is supported on recent GPU architectures, e.g., Nvidia Pascal and Volta architectures, to enable automatic data migration between device and host memories. For instance, when a device accesses a virtual page that is not mapped to a physical page on the device memory, a page fault is generated. Then, the runtime resolves the fault by remapping the page to a physical page on the device memory and copying the data. This procedure is also called on-demand paging. Now with the hardware-supported page fault and the runtime-managed data migration on UM, oversubscribing the GPU memory becomes feasible. For instance, when there is no physical memory available on the device for newly accessed pages, the runtime evicts pages from GPU to CPU and then bring the on-demand page. 

CUDA has introduced new features for optimizing the data migration on UM, i.e., \emph{memory advises} and \emph{prefetch}. Instead of solely relying on page faults, the \emph{memory advises} feature allows the programmer to provide data access pattern for each memory object so that the runtime can optimize migration decisions. The prefetch proactively triggers asynchronous data migration to GPU before the data is accessed, which reduces page faults and, consequently, the overhead in handling page faults. 

In this paper, we evaluate the effectiveness of these new memory features on CUDA applications using UM. Due to the absence of benchmarks designed for this purpose, we developed a benchmark suite of six GPU applications using UM. We evaluate the impact of the memory features in both in-memory and oversubscription executions on two platforms. The use of \emph{memory advises} results in performance improvement only when when we oversubscribe the GPU memory on the Intel-Volta/Pascal-PCI-E systems. On Power9-Volta-NVLink based system, using \emph{memory advises} leads to performance improvement only for in-memory executions. With GPU memory oversubscription, it results in substantial performance degradation. Our main contributions in this work are as follows:
\begin{itemize}
\item We survey state-of-art practice in UM \emph{memory advises}, \emph{prefetch}, and GPU memory oversubscription.
\item We design a UM benchmark suite consisting of six applications for evaluating advanced memory features. 
\item We evaluate the performance impact of \emph{memory advises}, \emph{prefetch} on two systems with Intel, Nvidia Pascal, and Volta GPUs connected via PCI-E and a system with IBM Power9 and Nvidia Volta GPU connected via NVLink.
\item Our results indicate that using \emph{memory advises} improves application performance in oversubscription execution on the Intel platform and in-memory executions on the IBM platform.
\item We show that UM \emph{prefetch} provides a significant performance improvement on the Intel-Volta/Pascal-PCI-E based systems while it does not show a performance improvement on the Power9-Volta-NVLink based system.
\end{itemize}

\section{Unified Memory}\label{sec:background}
In this section, we introduce the underlying mechanism in GPU UM, and the three memory advises. We also describe the prefetching and memory oversubscription.

\subsection{CUDA Unified Memory}
\begin{figure}[bt]
	\centering
		\includegraphics[width=0.8\linewidth]{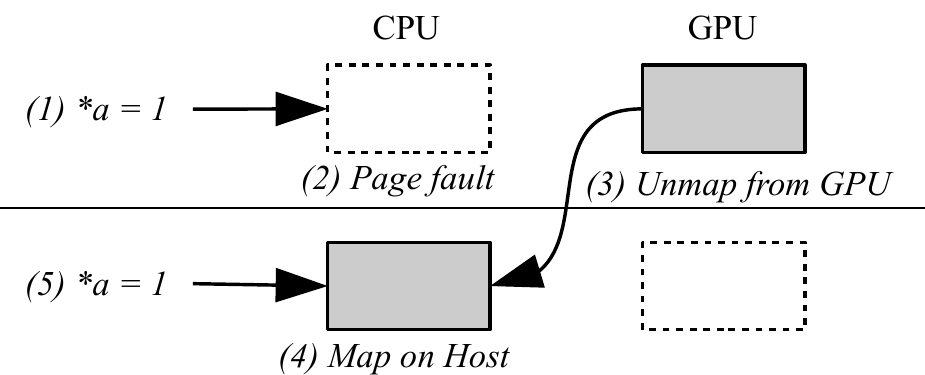}
		\caption{CPU writes to a page resident on the GPU, triggering a page fault and the page is migrated to CPU.\label{fig:page-fault}}
\end{figure}

UM creates a unified logical view of the physically separate memories across host and GPU. Currently, modern CPUs support 48-bit memory addresses while Unified Memory uses 49-bit virtual addressing, which can address both host and GPU memories~\cite{nvidia2016p100}. 
One of the main goals of Unified Memory is to provide a consistent view of data between devices. The system ensures a memory page can only be accessed by one process at a time. When a process accesses a page that is not resident of its memory system, a page fault occurs. The memory system holding the requested page will unmap it from its page table, and the page will be migrated to the faulting process. Figure~\ref{fig:page-fault} illustrates an example when the CPU accesses a page on GPU memory, and the page is migrated to CPU memory. Similarly, when GPU accesses a page not physically stored on GPU memory, the page will be moved to GPU.

UM was first introduced in CUDA 6.0~\cite{sakharnykh2018everything}.
Only until the recent Nvidia Pascal microarchitecture that has hardware support for page faults, bi-directional on-demand page migration becomes feasible~\cite{nvidia2016p100}. Resolving a page fault has high overhead, and memory thrashing that moves the same pages back and forth between the memories is even a performance bottleneck. The massive parallelism on GPU further exacerbates the page fault overhead because processes stall when page faults are being resolved, and multiple threads in different warps accessing the same page can cause multiple duplicated faults~\cite{Sakharnykh2017max}. 

\subsection{Data Movement Advises}
\begin{figure*}
	\centering
	\begin{subfigure}[b]{0.32\linewidth}
		\includegraphics[width=\linewidth]{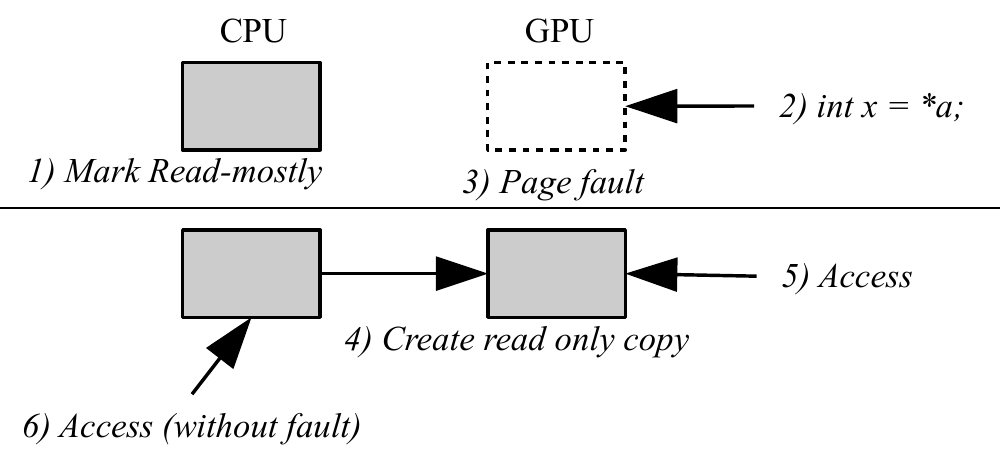}
		\caption{A read-mostly region is duplicated to the GPU to avoid page faults in the future.\label{fig:read-mostly}}
	\end{subfigure}
	~
	\begin{subfigure}[b]{0.3\linewidth}
		\includegraphics[width=\linewidth]{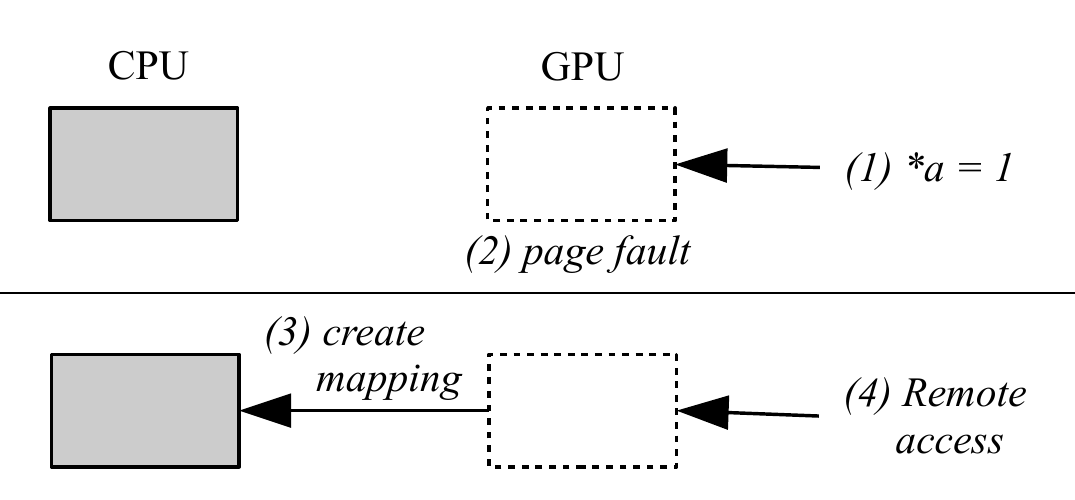}
		\caption{A host-preferred region is directly remote mapped to allow remote access from the GPU.\label{fig:preferred-location}}
	\end{subfigure}
	~
	\begin{subfigure}[b]{0.325\linewidth}
		\includegraphics[width=\linewidth]{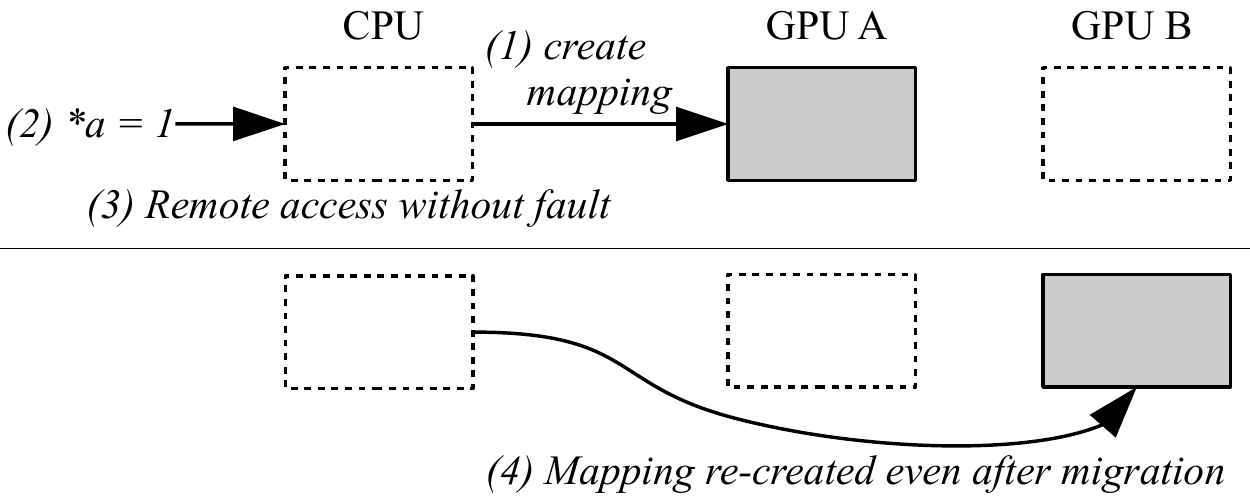}
		\caption{A GPU-resident region with accessed-by CPU advise can be accessed by CPU through remote memory access.\label{fig:accessed-by}}
	\end{subfigure}
	\caption{Page fault mechanism and effects of the three \emph{Memory Advise} in Unified Memory.}
\end{figure*}
CUDA 8.0 introduces a new programming interface, called memory advise~\cite{nvidia2019nvidia}. The concept is similar to {\tt posix\_madvise} in Linux, which uses application knowledge about access patterns to make informed decisions on page handling~\cite{posix}. The UM advise focuses on data locality, i.e., whether a page is likely to be accessed from the host or device. The main objective is to reduce unnecessary page migration and their associated overhead. Currently, developers can specify three access patterns to the CUDA runtime:

~\emph{cudaMemAdviseSetReadMostly} implies a read-intensive data region. In the basic UM, accessing a page on a remote side triggers page migration. However, with {\tt cudaMemAdviseSetReadMostly}, a read-only duplicate of the page will be created on the faulting side, which prevents page faults and data migration in the future. Figure~\ref{fig:read-mostly} illustrates an example, where the second access (step 5) has no page fault and is local access. This mechanism, however, results in a high overhead if there is any update to this memory region because all copies of the corresponding page will be invalidated to preserve consistency between different copies. Thus, this advice is often used in read-only data structures, such as lookup tables and application parameters.

~\emph{cudaMemAdviseSetPreferredLocation} sets the preferred physical location of pages. This advice pins a page and prevents it from migrating to other memories. Figure~\ref{fig:preferred-location} illustrates a page preferred on the host side, and GPU uses remote mapping to access the page. This advice established a direct (remote) mapping to the memory page. When accessing the page remotely, data is fetched through the remote memory instead of generating a page fault. If the underlying hardware does not support the remote mapping, the page will be migrated as in the standard UM. {\tt cudaMemAdviseSetPreferredLocation} is useful for applications with little data sharing between CPU and GPU, i.e., part of the application is executed completely on the GPU, and the rest of the application executes on the host. Data that is being used mostly by the GPU can be pinned to the GPU with the advice, avoiding memory thrashing.

~\emph{cudaMemAdviseSetAccessedBy} establishes a direct mapping of data to a specified device. Figure~\ref{fig:accessed-by} illustrates an example of a physical page on GPU being remotely access from the host. When \emph{cudaMemAdviseSetPreferredLocation} is applied, CUDA runtime tries to build a direct mapping to the page to avoid data migration so that the destination can access data remotely. Differently from~\emph{cudaMemAdviseSetPreferredLocation}, this \emph{cudaMemAdviseSetAccessedBy} does not try to pin pages on a specific device; instead, its main effect is to establish mapping on the remote device. This advice takes effect on the creation of the memory pages. The mapping will be re-established after the pages are migrated.

\subsection{Prefetching}
The CUDA interface introduces an asynchronous page prefetching mechanism, i.e., \emph{cudaMemPrefetchAsync()}~\cite{nvidia2019nvidia}, to trigger data migration. The data migration occurs in a background CUDA stream to avoid stalling the computation threads. One natural optimization for prefetching a large number of pages is to split into multiple streams, i.e., a bulk transfer, to prefetch pages in a batch of streams concurrently. If the page is prefetched to the device memory before the data access, no page faults will occur, and the GPU benefits from the high bandwidth on its local memory.

The behavior of the prefetching mechanism might change when used in combination with CUDA memory advises. For example, when \emph{cudaMemAdviseSetReadMostly} is set, a read-only copy will be immediately created. Also, when prefetching a region with \emph{cudaMemAdviseSetPreferredLocation} set to another destination memory, the pages will no longer be pinned to the preferred location. Thus, our evaluation considers the interplay between these two types of memory features. 

\subsection{Oversubscription of Device Memory}
GPU memory has a relatively small capacity compared to the system memory on CPU. One major limitation when porting large-scale applications to GPUs is to overcome their memory capacity to enable larger problems. UM in the post-Pascal page fault capable GPUs can oversubscribe GPU memory, allowing GPU kernels to use more memory than the physical capacity on the device. The memory oversubscription is achieved through the traditional virtual memory management, i.e., selected memory pages on the device are evicted to CPU to make space for newly requested pages. Currently, the CUDA runtime uses the Least Recently Used (LRU) replacement policy to select victim pages when running out of space~\cite{Sakharnykh2017uni}. Some work also proposed pre-eviction to start page eviction early to avoid stalling on the critical path~\cite{Ganguly:2019:IHP:3307650.3322224}.

\section{Methodology}\label{sec:methods}
We develop a benchmark suite for evaluating UM and different data migration policies. Although several porting efforts have been reported for specific applications, there lacks a suite of diverse kernels for controlled experiments across platforms. Thus, we extend the memory management in popular GPU benchmark and applications to utilize UM with advanced advise and prefetching features.

\subsection{Application and Benchmarks}

\begin{table*}
	\caption{Applications and data input sizes on different platforms.\label{table:app}}
	\resizebox{\textwidth}{!}{%
		\begin{tabular}{|l|l|l|l|l|l|}
			\hline
			\multicolumn{1}{|c|}{\multirow{2}{*}{Name}} & \multicolumn{1}{c|}{\multirow{2}{*}{Description}} & \multicolumn{2}{c|}{\begin{tabular}[c]{@{}c@{}}Input size Intel-Pascal\\ (Approximate GB)\end{tabular}} & \multicolumn{2}{c|}{\begin{tabular}[c]{@{}c@{}}Input size Intel-Volta \& P9-Volta\\ (Approximate GB)\end{tabular}} \\ \cline{3-6} 
			\multicolumn{1}{|c|}{} & \multicolumn{1}{c|}{} & \multicolumn{1}{c|}{In-memory} & \multicolumn{1}{c|}{Oversubscribe} & \multicolumn{1}{c|}{In-memory} & \multicolumn{1}{c|}{Oversubscribe} \\ \hline
			Black-Scholes (BS) & A financial application that performs option pricing. & 4 & 6.4 & 15.2 & 26 \\
			Matrix Multiplication (cuBLAS) & A general matrix matrix multiplication in single precision using cuBLAS. & 3.9 & 6.3 & 15.2 & 25.4 \\ 
			Conjugate Gradient (CG) & A conjugate gradient solver that solves a sparse linear system using cusparse. & 3.8 & 6.4 & 15.4 & 25.4 \\
			Graph500 & Breadth-first search (BFS) kernel of Graph500. & 3.63 & 7.62 & 8.52 & N/A \\
			Convolution 0 (conv0) & A FFT-based image convolution using Real-to-Complex and Complex-to-Real FFT plans. & 2.8 & 6.4 & 11.6 & 25.6 \\
			Convolution 1 (conv1) & A FFT-based image convolution using Complex-to-Complex FFT plan. & 3.5 & 6.7 & 13.6 & 25.5 \\
			Convolution 2 (conv2) & A FFT-based image convolution using Complex-to-Complex FFT plan. & 3.0 & 6.4 & 11.6 & 25.5 \\
			Finite Difference Time Domain (FDTD3d) & A finite difference solver in three dimension. & 3.8 & 6.4 & 15.2 & 25.3 \\ \hline
		\end{tabular}%
	}
\end{table*}

Our benchmark suite includes six applications, as specified in Table~\ref{table:app}. These applications include numerical solvers, financial application, image processing, and graph problems. The benchmark suite is available at  a repository \footnote{\url{https://github.com/steven-chien/um-apps}}. 

For each application, we develop four versions in addition to the original version that uses explicit GPU memory allocation. Our benchmarks use long data types to support large input problems in oversubscription executions. We use GPU kernel execution time as the figure of merit. 

We present detailed tracing results for BS, CG, and FDTD3d on selected platforms to study the implications of data movement. BS is a financial application that performs option pricing. BS features good data reuse because the same input data set is used in multiple iterations in the application lifetime. CG is a linear solver that solves a linear system $Ax=b$ on the GPU. An error is computed on the host using the results from GPU computation after the solving iteration finishes. FDTD3d is a finite difference solver that reads and writes to two arrays in an interleaving manner. Both arrays are being initialized using the same data. The output eventually resides in one of the arrays.

\subsubsection{UM} The first version is an implementation that uses UM with minimal changes. We simply replace the memory allocation in applications from \emph{cudaMalloc()} to \emph{cudaMallocManaged()} and eliminate explicit data copy, i.e., \emph{cudaMemcpy()}, between host and device. After the completion of a GPU kernel, if the application has no subsequent host computation using the GPU results, an explicit data copy by \emph{memcpy()} is inserted to simulate a CPU computation using the results.

\subsubsection{UM Advise} The second version is UM with Advise. This version is based on the basic UM version and applies memory advises to data structures in the application. A stall in GPU execution, e.g., for resolving page fault, has a significant impact on performance due to massive parallelism. Thus, the main consideration for memory advises is to keep data used by GPU close to GPU memory. Therefore, we set a \emph{cudaMemAdviseSetPreferredLocation} and specify the preferred location to GPU memory after the memory allocation of a data structure that is accessed by GPU in the computation. If the data structure is initialized by the CPU, we set a \emph{cudaMemAdviseSetAccessedBy} CPU to keep the data physically on GPU but establish a remote mapping on CPU. With this optimization, the host data initialization performs remote accesses to initialize data in GPU memory directly. For constant data structures, the \emph{cudaMemAdviseSetReadMostly} advice is set after data initialization. This optimization will only have page fault at the first access but keep all subsequent accesses local. 

\subsubsection{UM Prefetch} The third version is UM with prefetch. We apply {\tt cudaMemPrefetchAsync} to trigger page migration at appropriate sites explicitly. We prefetch large data structures that will be accessed by GPU kernels in a background stream while the GPU kernel is launched in the default stream. After completing the GPU kernel execution, we prefetch the arrays containing results to the host memory in the default stream. One advantage of bulk transfer in prefetch, compared to resolving individual page fault groups, is high memory bandwidth to utilize the hardware capability fully. Explicitly triggering page pages in bulk improves transfer efficiency. Furthermore, to prefetch pages avoids page faults as data already resides in the physical memory when the kernel starts executing.

\subsubsection{UM Both} Finally, in the fourth version, we combine memory advises and prefetch to examine the mutual effects of both techniques.

\subsection{Test Environment}
We evaluate our benchmark applications on three platforms:
\begin{enumerate}
	\item {\bf Intel-Pascal} is a single node system with Intel Core i7-7820X processor and 32 GB of RAM. It has one GeForce GTX 1050 ti GPU with 4GB memory. The GPU is connected through PCIe. The operating system is Ubuntu 18.10 and the host compiler is GCC 8.3.
	\item {\bf Intel-Volta} is a GPU node on Kebnekaise at HPC2N in Umeå. It has an Intel Xeon Gold 6132 processor with 192 GB of RAM. The node has two Tesla V100 GPU with 16 GB memory and the GPU is connected through PCIe. The operating system is Ubuntu 16.04 and the host compiler is GCC 8.2.
	\item {\bf P9-Volta} is a node with an IBM Power9 processor and 256 GB of RAM. The system has four Tesla V100 GPUs with 16 GB of HBM. The GPU is connected through NVLINK to CPU.
\end{enumerate}

Our platforms consist of two Intel systems that use Pascal and Volta GPUs, and a Power9 system that uses Volta GPU. All the systems use CUDA 10.1. We only use one GPU in the experiments. For each application variation, we perform benchmark runs up to five times and present the average GPU kernel execution time and standard deviation. An exception is Graph500, where we report the average and standard deviation of BFS iterations. We separate our experiments into two cases: when problem size fits into GPU memory and when oversubscription of memory is required. Their problem sizes are selected to be approximately 80\% and 150\% to GPU memory, respectively. A detailed list of sizes is presented in Table~\ref{table:app}. Due to the limitation in implementation for input data size, we only examine Graph500 with oversubscription on Intel-Pascal. However, the input size does not follow the 150\% data size rule.

Apart from benchmark executions, we perform profiling runs using \emph{nvprof} for selected applications. We obtain the trace by \emph{--print-gpu-trace}. By selecting entries with \emph{Unified Memory Memcpy HtoD} and \emph{Unified Memory Memcpy DtoH}, we can build a time series of data movement. Through a comparison of the time series and time spent on memory movement, it is possible to compare and characterize the intensity of data movement between different application variations. 
\section{Results}\label{sec:results}
In this section, we present the performance and profiling results of the applications in four configurations: basic UM (UM), UM with Advise (UM Advise), UM with Prefetch (UM Prefetch) and UM with both Advise and Prefetch (UM Both). Each application is evaluated in each configuration with two problem sizes: one that fits into GPU memory (in-memory execution) and one that oversubscribes GPU memory (oversubscription execution). We report the average and standard deviation of GPU kernel execution time for each application.



\subsection{In-Memory Execution}\label{sec:result-in-memory}

\begin{figure*}[t]
	\centering
	\begin{subfigure}[b]{0.322\textwidth}
		\includegraphics[width=\linewidth]{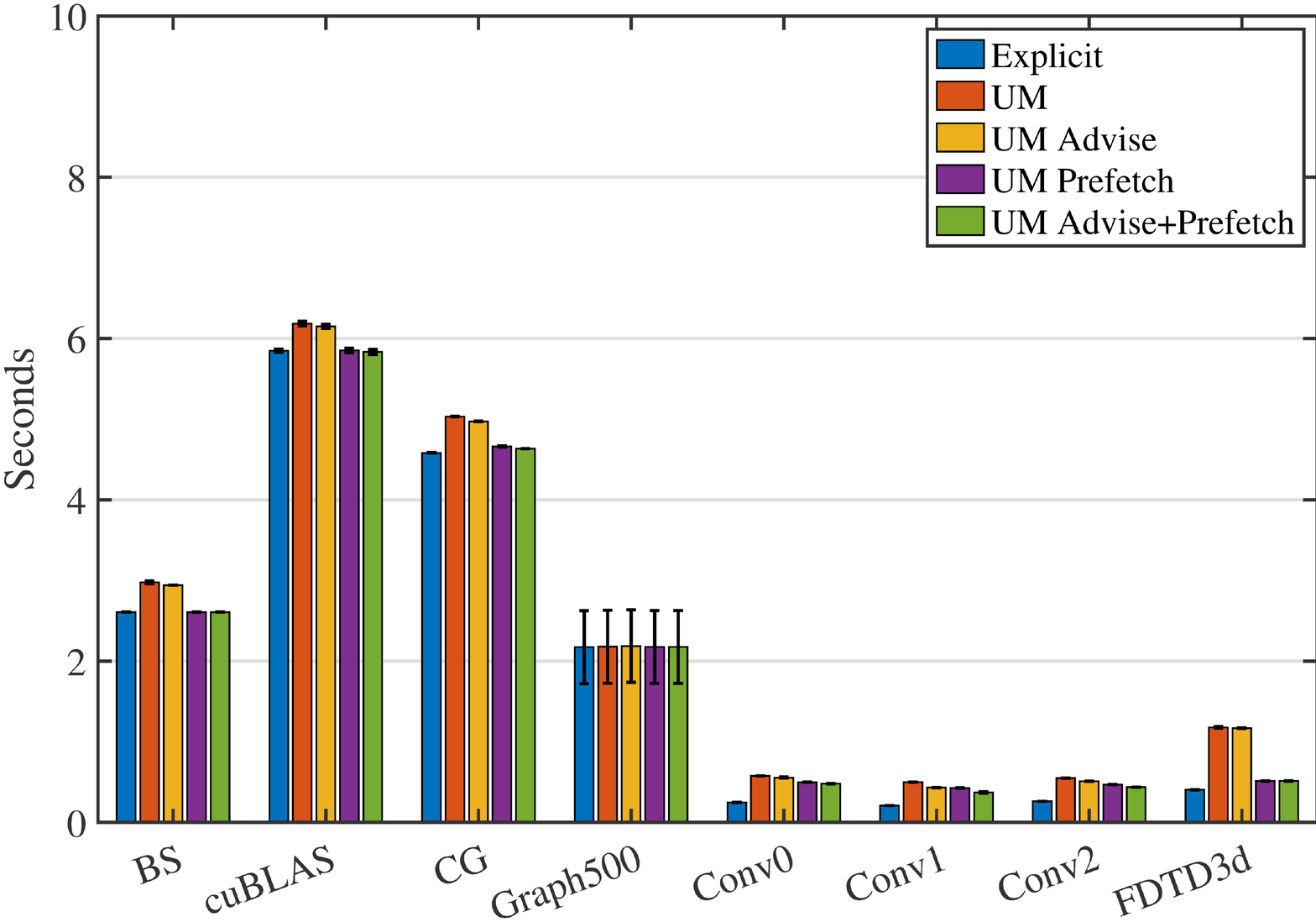}
		\caption{Intel-Pascal}
	\end{subfigure}
	~
	\begin{subfigure}[b]{0.322\textwidth}
		\includegraphics[width=\linewidth]{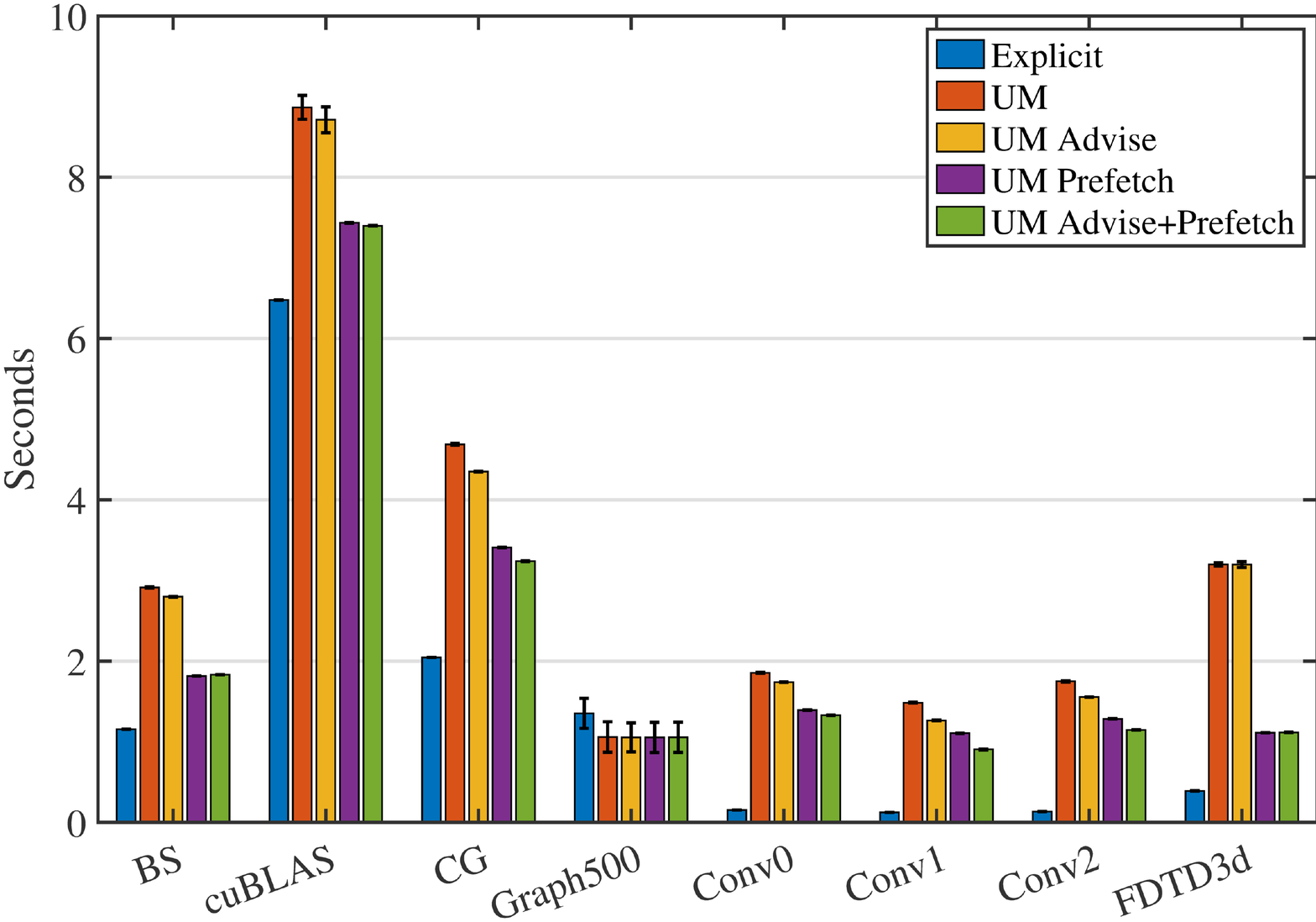}
		\caption{Intel-Volta}
	\end{subfigure}
	~
	\begin{subfigure}[b]{0.32\textwidth}
		\includegraphics[width=\linewidth]{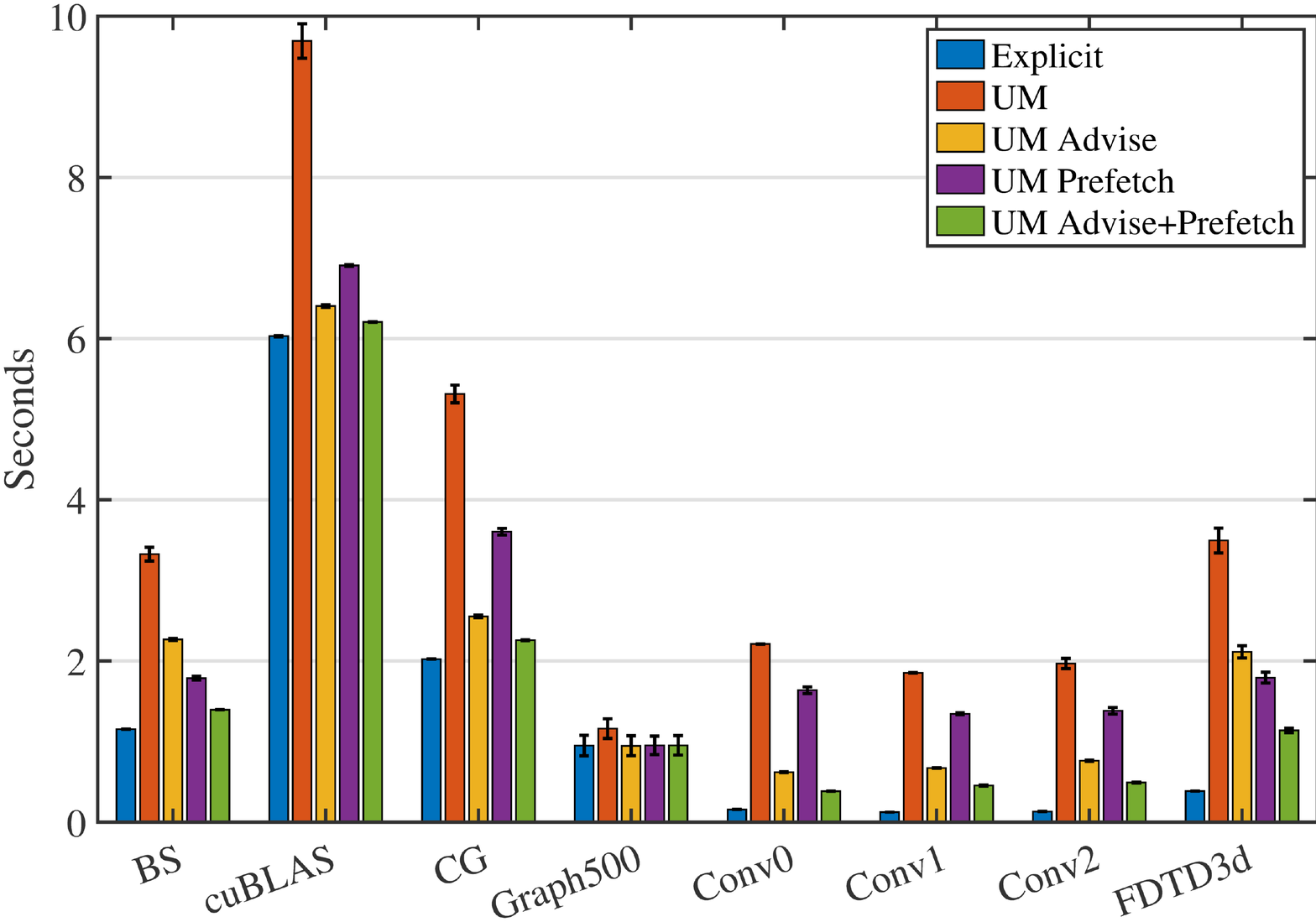}
		\caption{Power9-Volta.}
	\end{subfigure}
	\caption{GPU kernel execution time of applications where data fits in GPU memory.}
	\label{fig:gpu-time-in-memory}
\end{figure*}

\begin{figure*}[t]
	\centering
	\begin{subfigure}[b]{0.2\textwidth}
		\includegraphics[width=\textwidth]{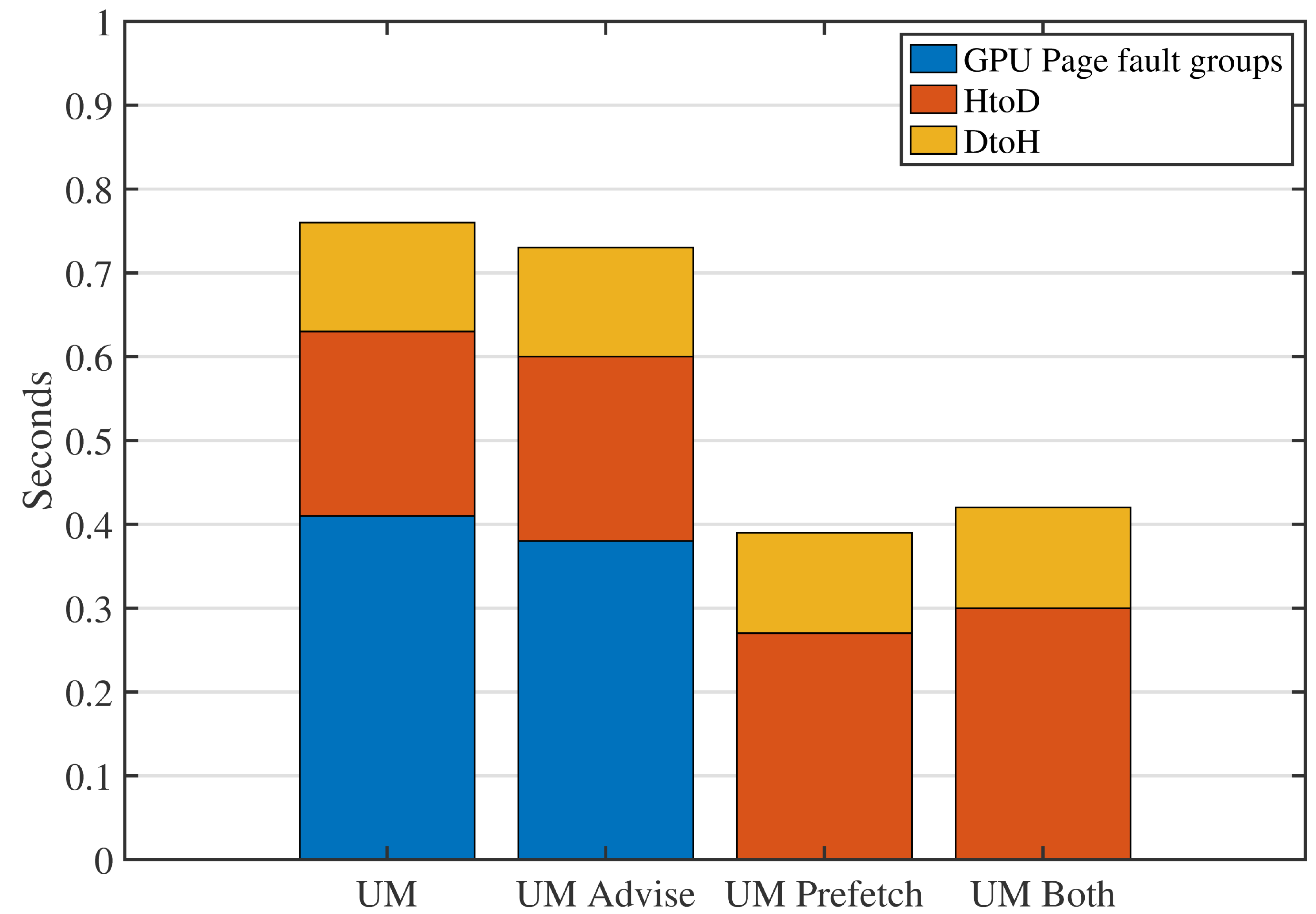}
		\caption{BS on Intel-Pascal}
		\label{fig:in-memory-pagefault-breakdown-pascal-BS}
	\end{subfigure}
	~
	\begin{subfigure}[b]{0.2\textwidth}
		\includegraphics[width=\textwidth]{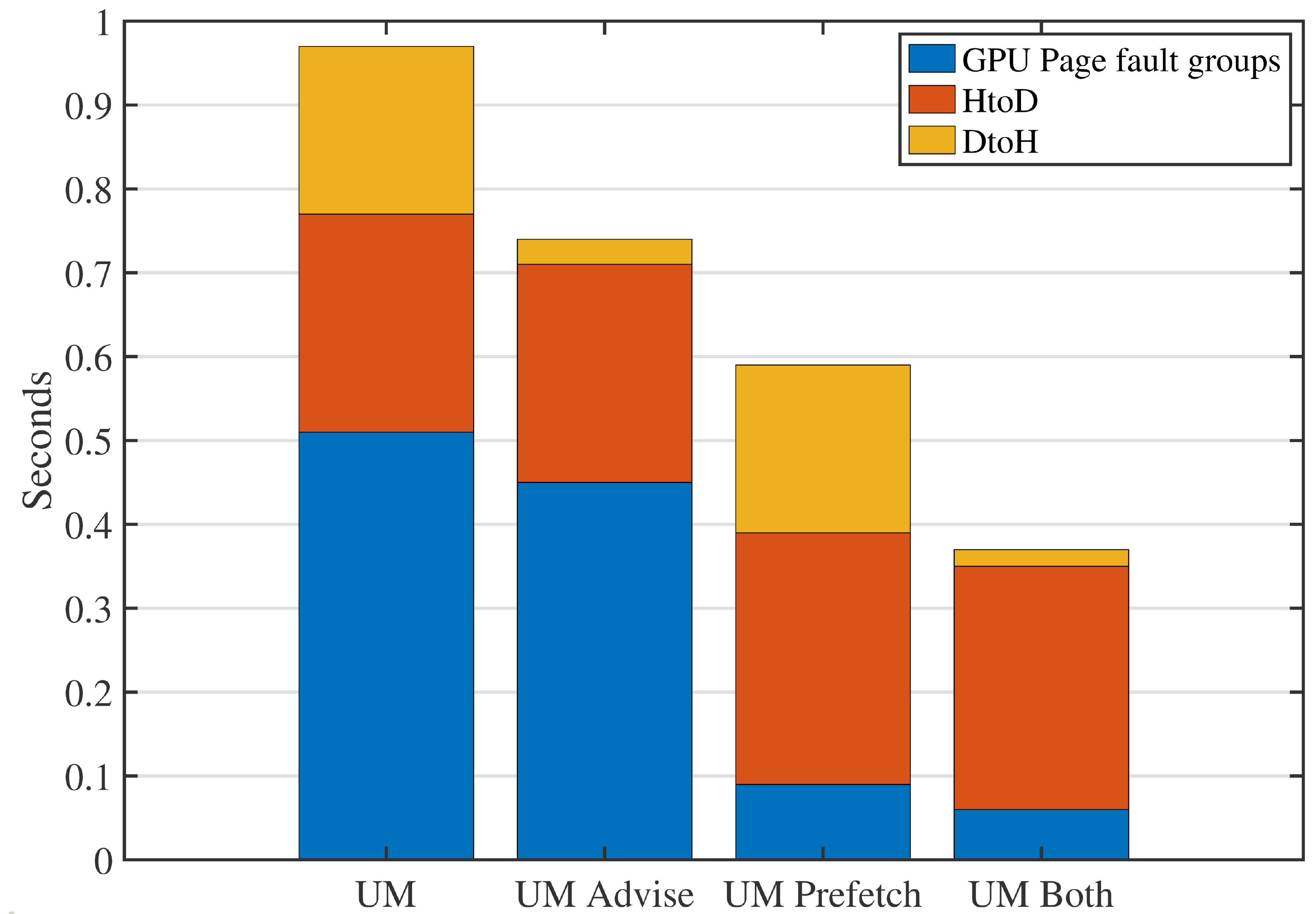}
		\caption{CG on Intel-Pascal}
		\label{fig:in-memory-pagefault-breakdown-pascal-CG}
	\end{subfigure}
	~
	\begin{subfigure}[b]{0.2\textwidth}
		\includegraphics[width=\textwidth]{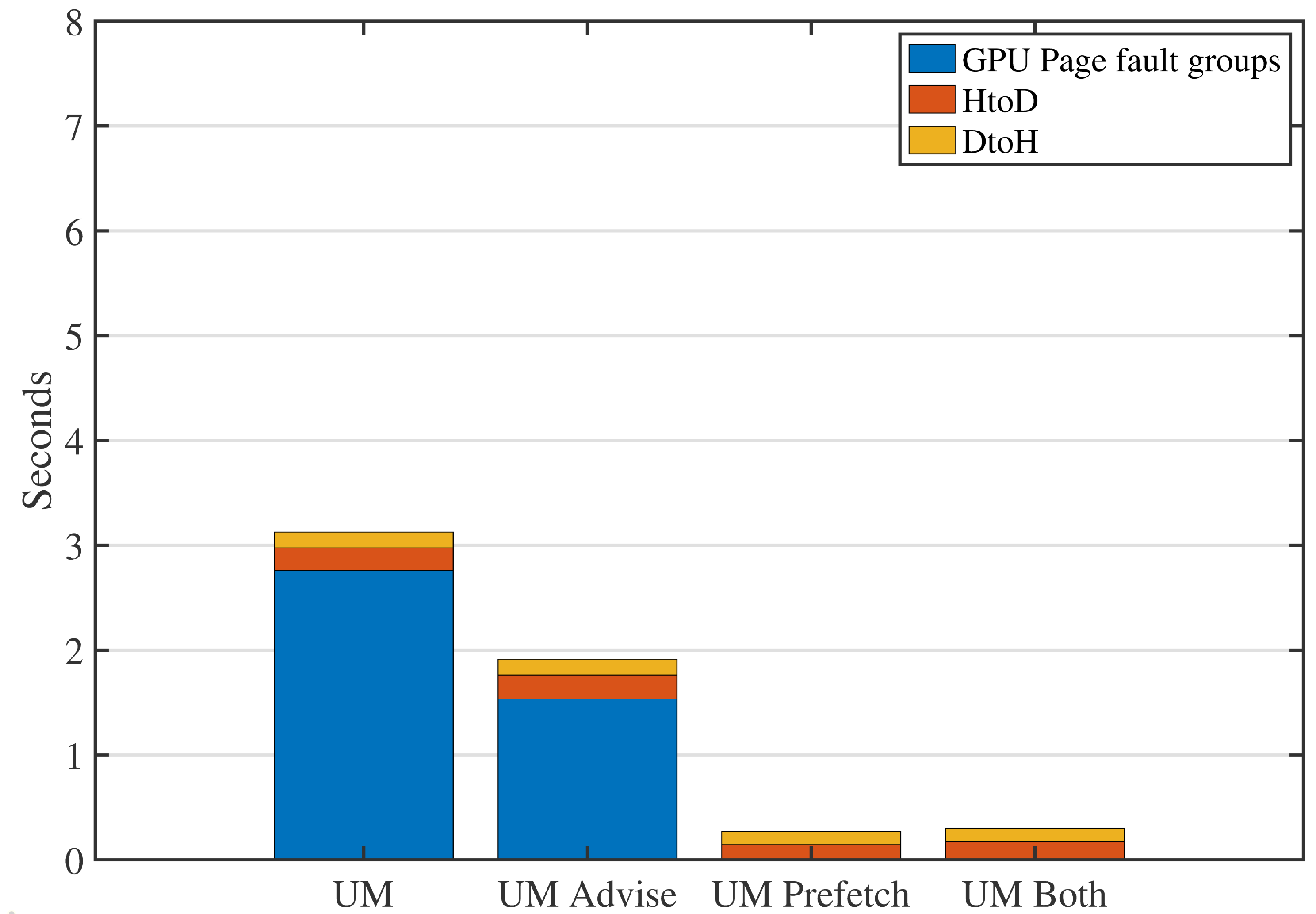}
		\caption{BS on P9-Volta}
		\label{fig:in-memory-pagefault-breakdown-p9-BS}
	\end{subfigure}
	~
	\begin{subfigure}[b]{0.2\textwidth}
		\includegraphics[width=\textwidth]{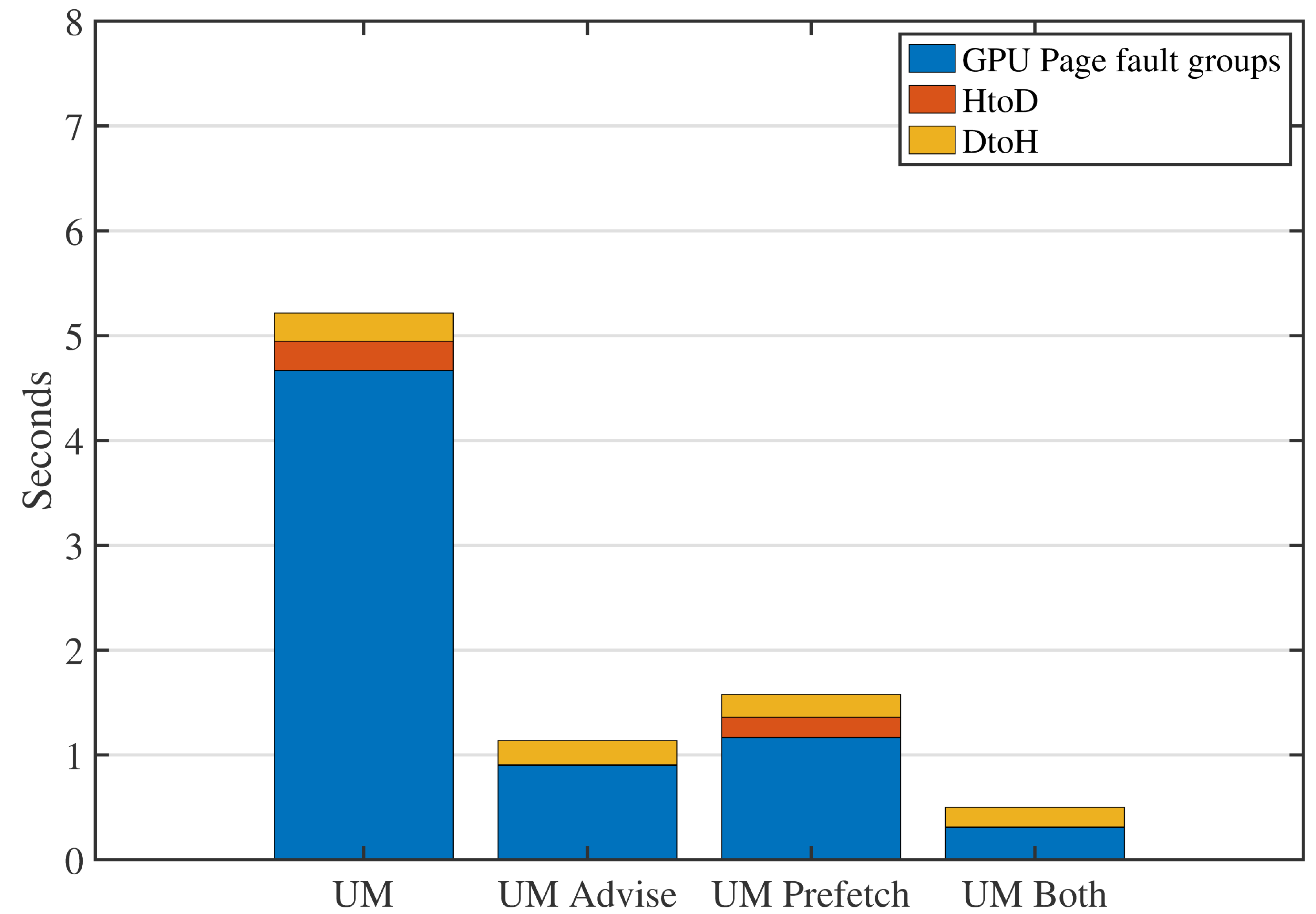}
		\caption{CG on P9-Volta}
		\label{fig:in-memory-pagefault-breakdown-p9-CG}
	\end{subfigure}
	\caption{Breakdown of total time spent handling page faults and data movement when applications are running in-memory.}
	\label{fig:pagefault-breakdown-in-memory}
\end{figure*}

\begin{figure*}[t]
	\centering
	\begin{subfigure}[b]{0.225\textwidth}
		\includegraphics[width=\textwidth]{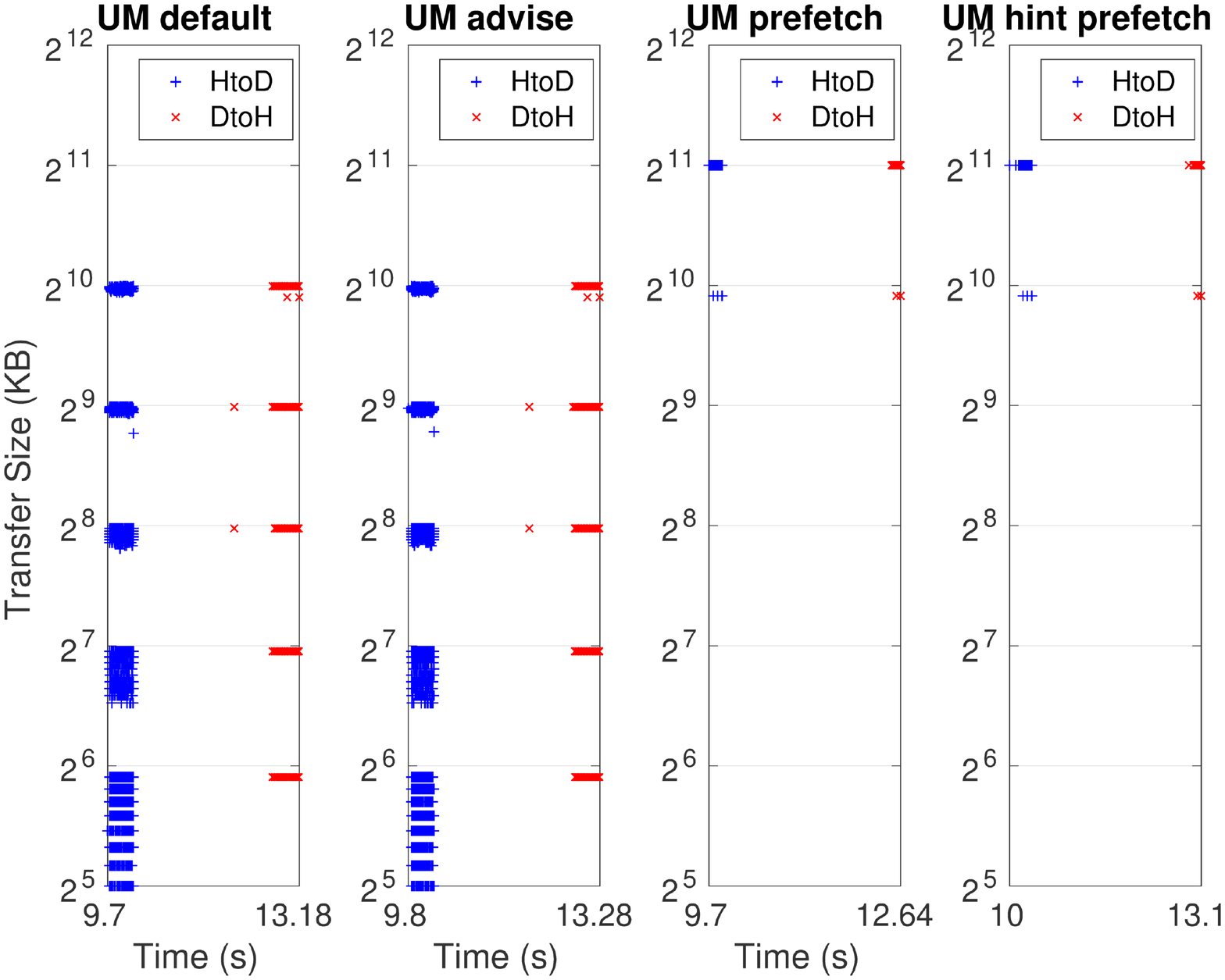}
		\caption{BS Intel-Pascal}
		\label{fig:trace-in-memory-pascal-BS}
	\end{subfigure}
	~
	\begin{subfigure}[b]{0.225\textwidth}
		\includegraphics[width=\textwidth]{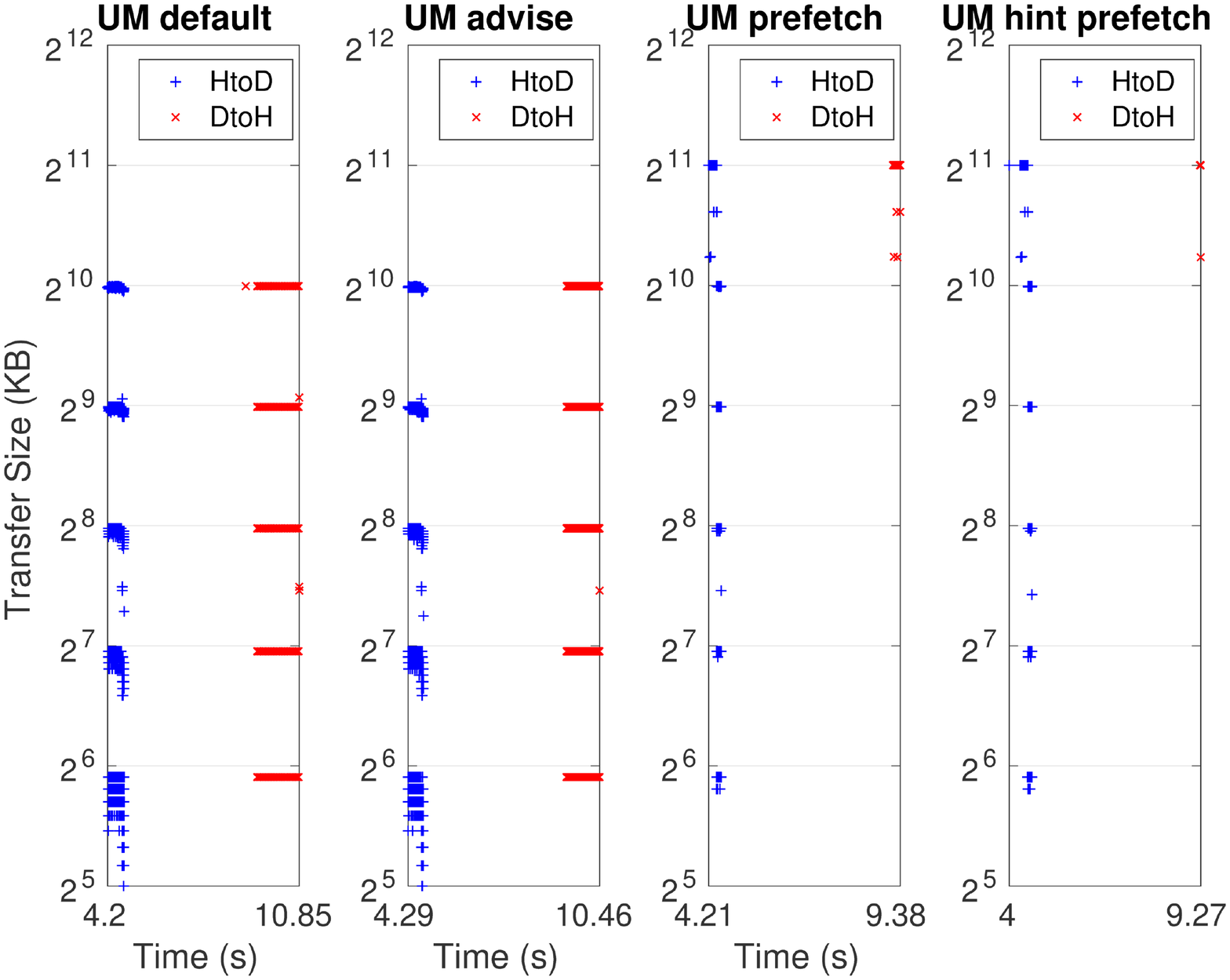}
		\caption{CG Intel-Pascal}
		\label{fig:trace-in-memory-pascal-CG}
	\end{subfigure}
	~
	\begin{subfigure}[b]{0.225\textwidth}
		\includegraphics[width=\textwidth]{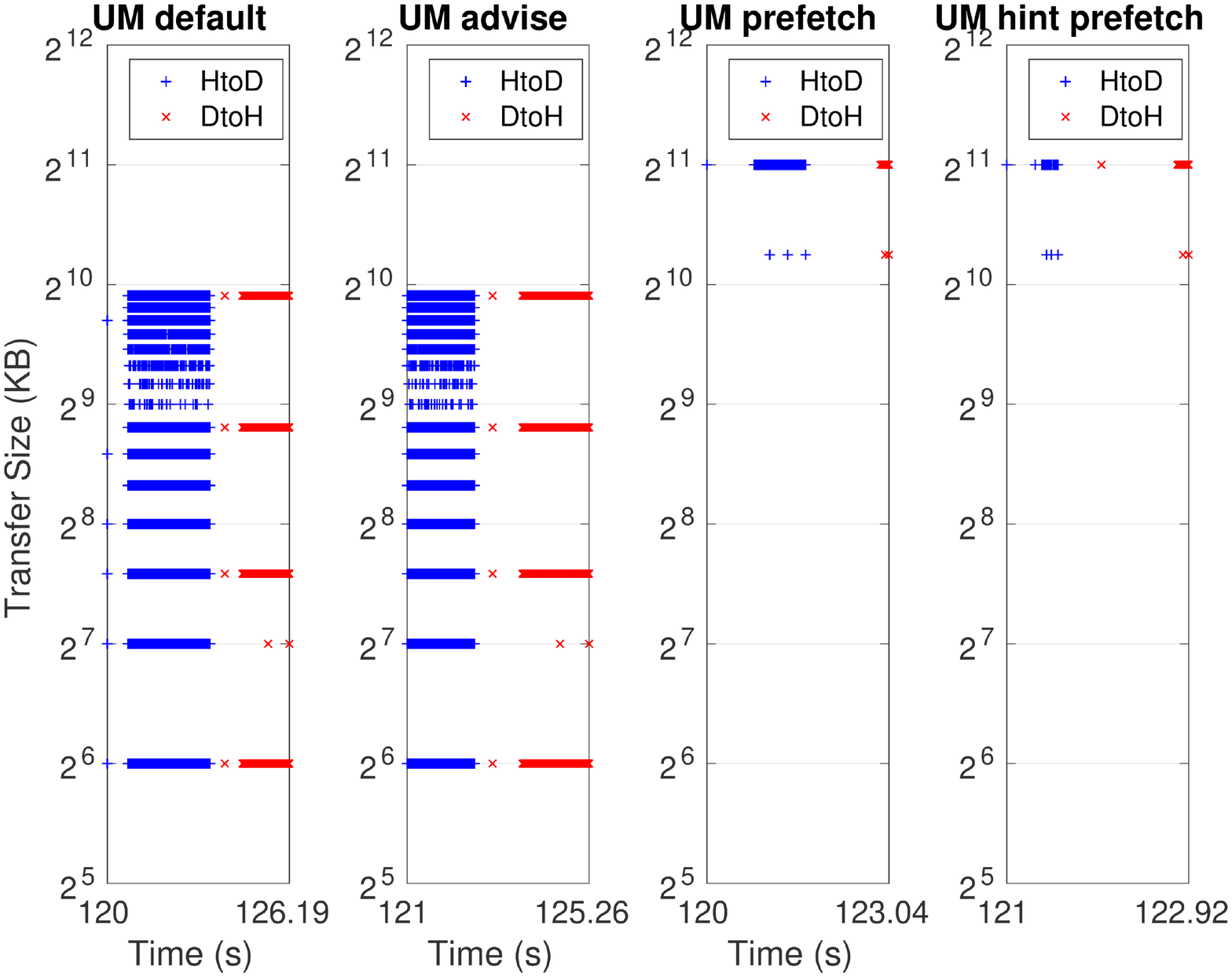}
		\caption{BS P9-Volta}
		\label{fig:trace-in-memory-p9-BS}
	\end{subfigure}
	~
	\begin{subfigure}[b]{0.225\textwidth}
		\includegraphics[width=\textwidth]{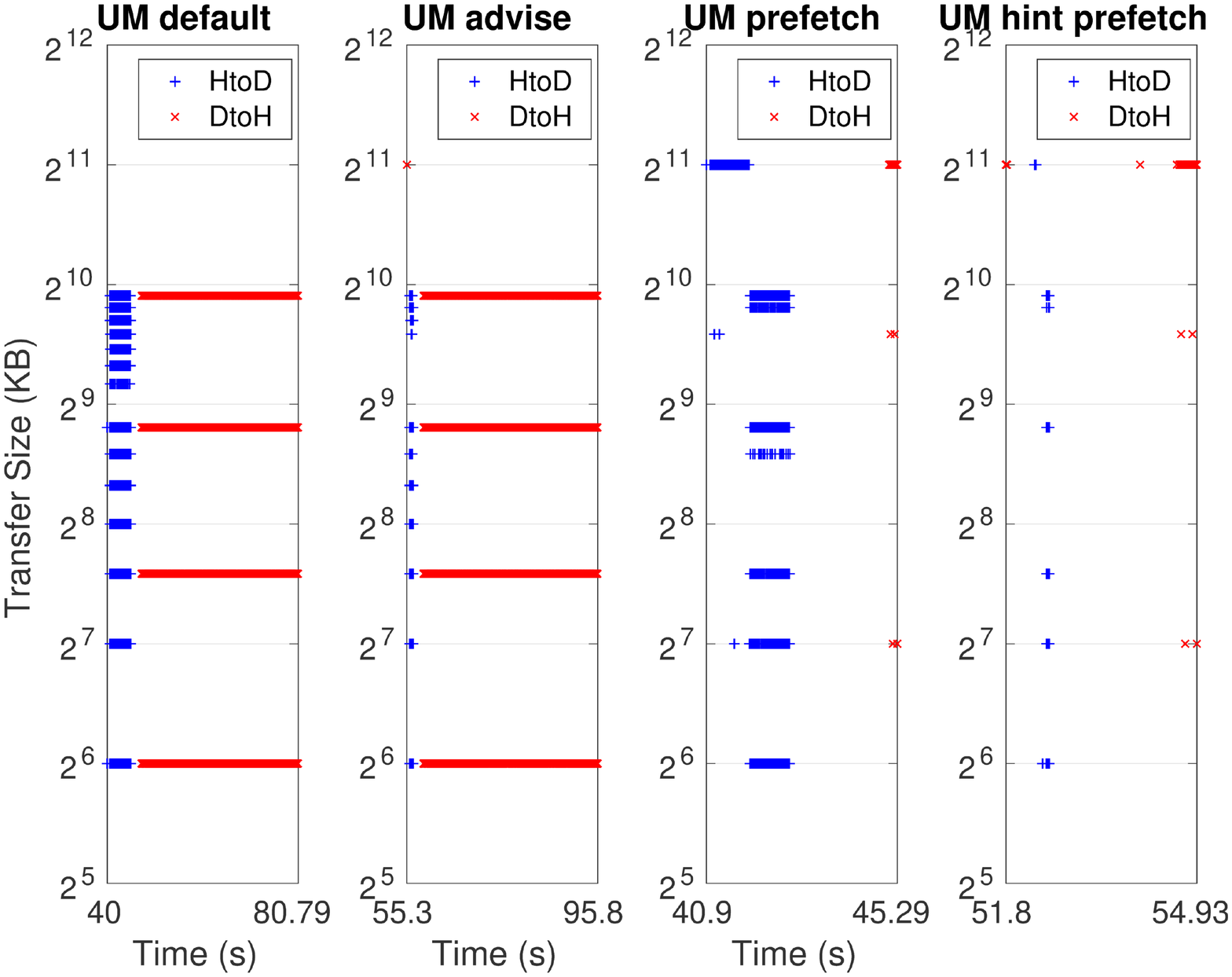}
		\caption{CG P9-Volta}
		\label{fig:trace-in-memory-p9-CG}
	\end{subfigure}
	\caption{UM data transfer traces when running in-memory.}
	\label{fig:trace-in-memory}
\end{figure*}

We present the GPU kernel execution time of the applications in Fig.~\ref{fig:gpu-time-in-memory}. The performance of all applications decreases when using basic UM instead of explicit data movement between CPU and GPU memories. Performances on Volta GPUs platforms have a larger performance decrease. In particular, our convolution and FDTD3d applications exhibit a drastic increase in execution time. The execution time of conv2 and FDTD3d are $14\times$ and $9\times$ higher respectively on P9-Volta. Performance change is similar to Intel-Volta. Performance decrease is less drastic but still considerable on Intel-Pascal. The execution of both applications is $2-3\times$ slower than the execution time of applications using explicit data movement.


After applying advises, the performance of our applications generally improves. It is possible to improve execution time up to 15\% on Intel-based platforms. The impact of advises is higher for the three FFT based convolution applications on Intel platforms. Advises have a significant impact on all the applications, and execution time can be improved by up to 70\% on the P9 platform. Applications, such as CG and cuBLAS, results in similar execution time to the original version with explicit memory allocation.  This implies that some advises are more effective than others on the P9 platform.

Expensive page fault handling can be avoided by prefetching data to the GPU before execution. Our results show that prefetch has a much higher impact on Intel-based platforms than P9-based platforms. Application performance generally improves when prefetch is used: our results show that it has a much higher impact on Intel-based platforms than the case advise is used. The performance of FDTD3d improves by up to 56\% on the Intel-Pascal system. The performance of Black-Scholes application is close to on-par with the application version using explicit data transfer. As for Intel-Volta, the performance of FDTD3d improves by up to 65\%. Performance improves by 50\% on the P9-Volta system. However, the improvement is less than when only advises are applied. Despite that, we notice that when both advises and prefetch are used together, it generally outperforms the performance of applications using only advises or prefetch.

To better understand the difference in terms of data movement between the versions, we plot the total time spent on different UM events in Fig.~\ref{fig:pagefault-breakdown-in-memory} as stacked bar plots. They show the total time spent on GPU page fault group handling and data transfer, respectively. In particular, we have selected BS and CG for the comparison. The bar plot reveals two important information for comparison: the time spent on data movement, which correlates to the amount and efficiency of data transferred, as well as stalls due to page fault, which correlates to the number of page fault and efficiency of fault resolution.

Since the Black-Scholes application uses the same input dataset repeatedly over iterations, when data size fits in memory, the first iteration will be slower due to page migration. Subsequent iterations should be able to execute at full speed as data already resides in device memory. For this particular application, the advise \emph{cudaMemAdviseSetReadMostly} is applied to the input arrays. No other advise is applied. The same goes for prefetch. Figs.~\ref{fig:in-memory-pagefault-breakdown-pascal-BS} and~\ref{fig:in-memory-pagefault-breakdown-p9-BS} show the break down of total time spent on data-related activities on the two platforms for the Black-Scholes application. Comparing to Intel-Pascal, the data transfer is much faster on P9-Volta, while the impact of stalling is less profound on Intel-Pascal. This can be attributed to the larger input data used and a faster interconnect on P9-Volta. For UM Advise, the time spent on data transfer is similar while the time spent stalling due to page fault has reduced. This suggests that page fault handling becomes more efficient when the advises are applied. The observation is similar for both Intel-Pascal and Intel-Volta similarly. When prefetch is used, the same amount of data is being transferred while the stall due to page fault is eliminated. This implies the complete elimination of page faults. By prefetching pages in bulk, data can be transferred at a fast pace to avoid future page faults when accessed. The observation can be confirmed by Figs.~\ref{fig:trace-in-memory-pascal-BS} and~\ref{fig:trace-in-memory-p9-BS}, where the detailed transfers are plotted as a time series. When prefetch is applied, data is transferred as a block at a much higher rate.

The Conjugate Gradient application solves a linear system $Ax=b$ iteratively. When applying advises, we set the preferred location of matrix $A$ and vector $b$ to GPU memory. We also set a read-mostly advise on the sparse matrix after completing initialization. The breakdown of time spent on Intel-Pascal and P9-Volta are shown in Figs.\ref{fig:in-memory-pagefault-breakdown-pascal-CG} and~\ref{fig:in-memory-pagefault-breakdown-p9-CG}, respectively. The use of advises results in similar time spent on data transfer from host to device but a slight reduction in time on stalls on Intel-Pascal system. A considerable reduction in time spent on the host to device transfer and stall is observed on the P9-Volta system. One reason is the use of preferred location advise, where the data arrays are initialized from the host on GPU memory through remote memory access. On Power9, it is possible for the CPU to access GPU memory while this is not possible on Intel platforms. At the same time, time spent on transfer from device to host is largely eliminated on Intel-Pascal. One possible reason is due to the read-mostly advise. Instead of migrating pages to the GPU from host memory, a read-only copy is copied to the GPU. This means that a copy of data exists in both memory systems. When the $Ax$ is being computed, $A$ can be fetched directly in host memory. Since P9-Volta initializes data directly in GPU memory, a copy has to be fetched back to the host. In this case, the naive use of prefetch results in a reduction of time spent stalling. Despite the fact that more data is transferred from device to host, the use of prefetch results in a higher transfer rate. The data transfer trace is presented in Figs.~\ref{fig:trace-in-memory-pascal-CG} and~\ref{fig:trace-in-memory-p9-CG}. When used in combination with advises, it results in a reduction of time for data transfer and stall.

\subsection{Oversubscription Execution}\label{sec:result-oversubscribe}

\begin{figure*}[t]
	\centering
	\begin{subfigure}[b]{0.322\textwidth}
		\includegraphics[width=\linewidth]{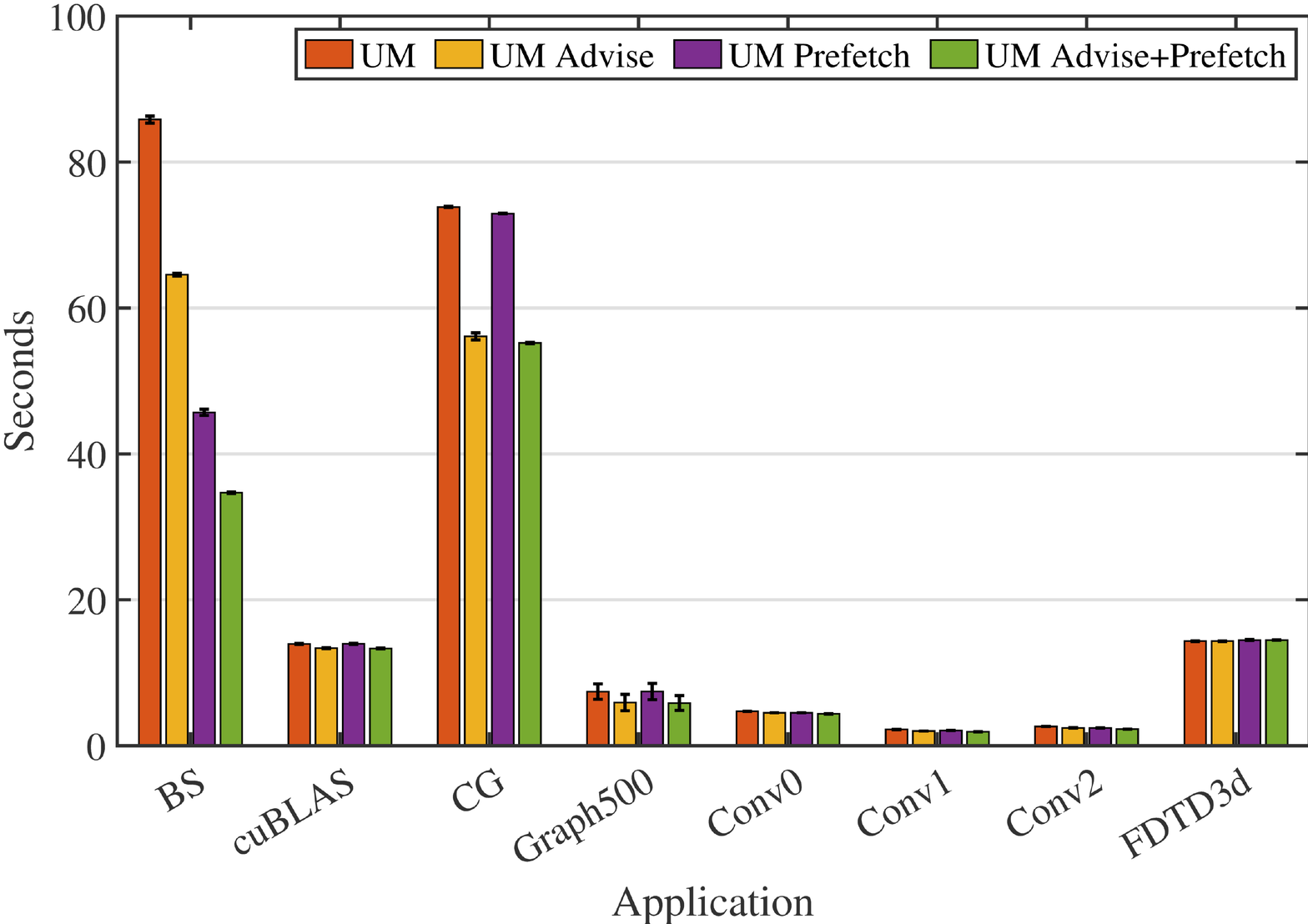}
		\caption{Intel-Pascal}
	\end{subfigure}
	~
	\begin{subfigure}[b]{0.322\textwidth}
		\includegraphics[width=\linewidth]{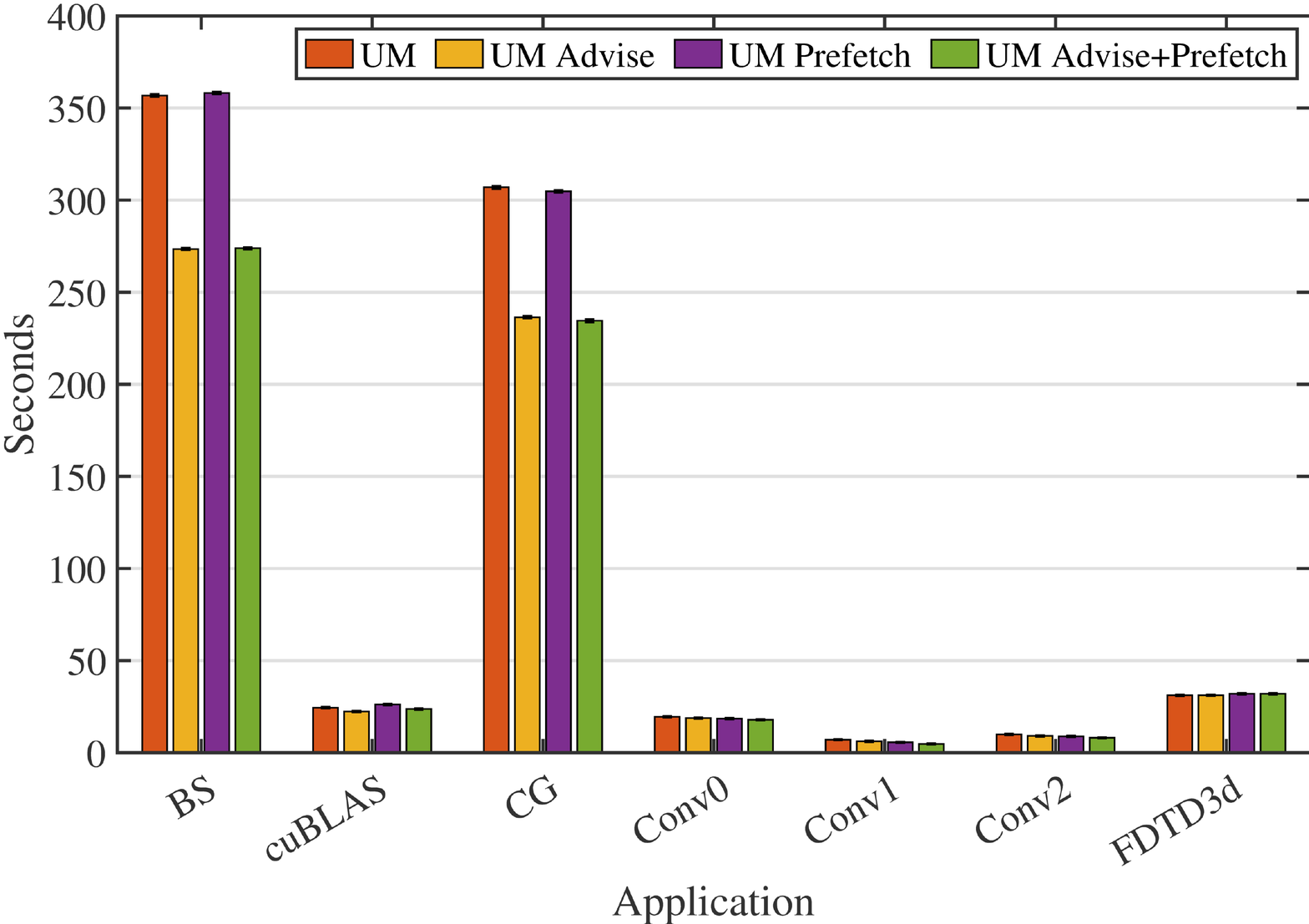}
		\caption{Intel-Volta}
	\end{subfigure}
	~
	\begin{subfigure}[b]{0.32\textwidth}
		\includegraphics[width=\linewidth]{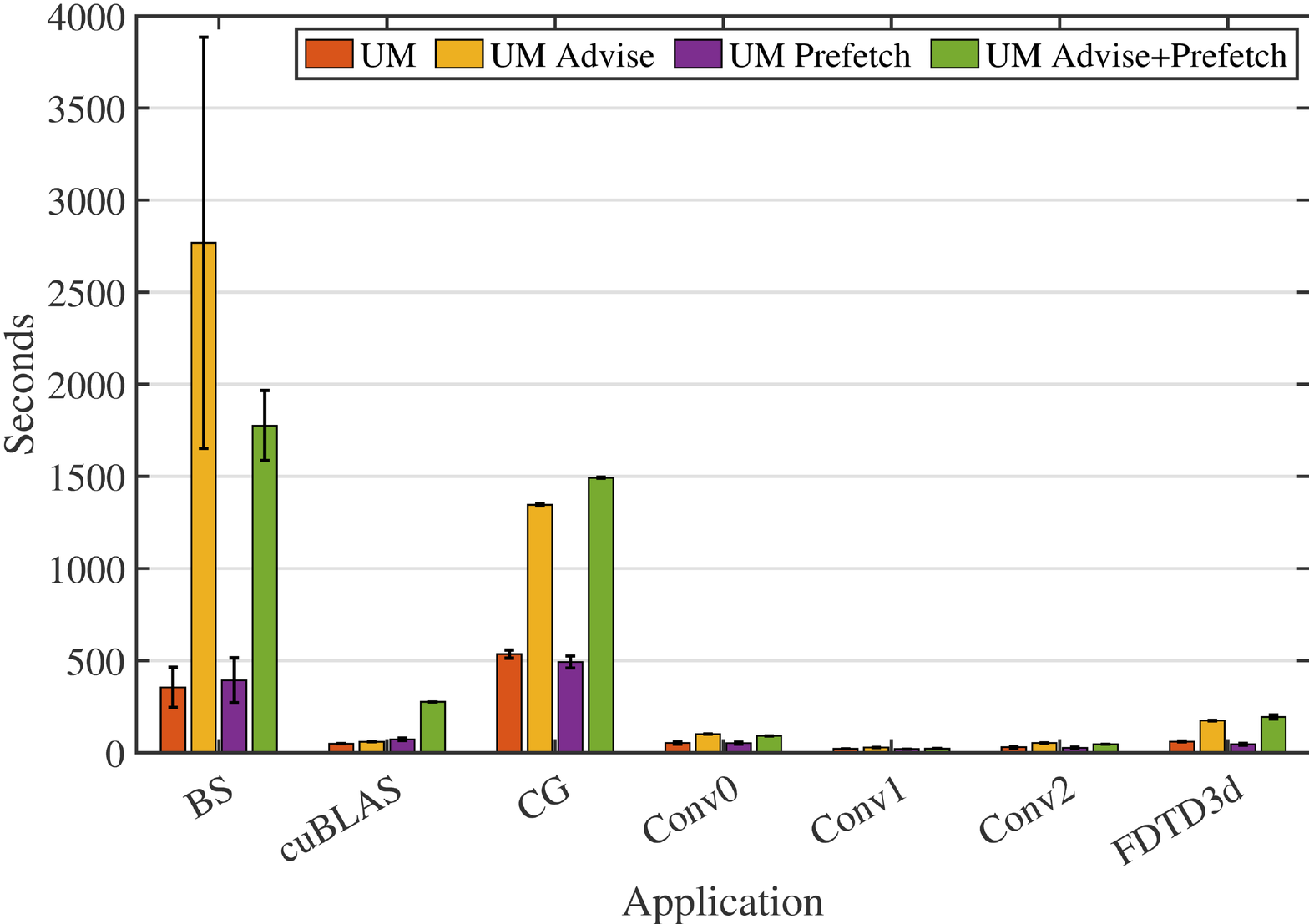}
		\caption{Power9-Volta.}
	\end{subfigure}
	\caption{GPU kernel execution time of applications where data do not fit in GPU memory.}
	\label{fig:gpu-time-oversubscribe}
\end{figure*}

\begin{figure*}
	\centering
	\begin{subfigure}[b]{0.2\textwidth}
		\includegraphics[width=\textwidth]{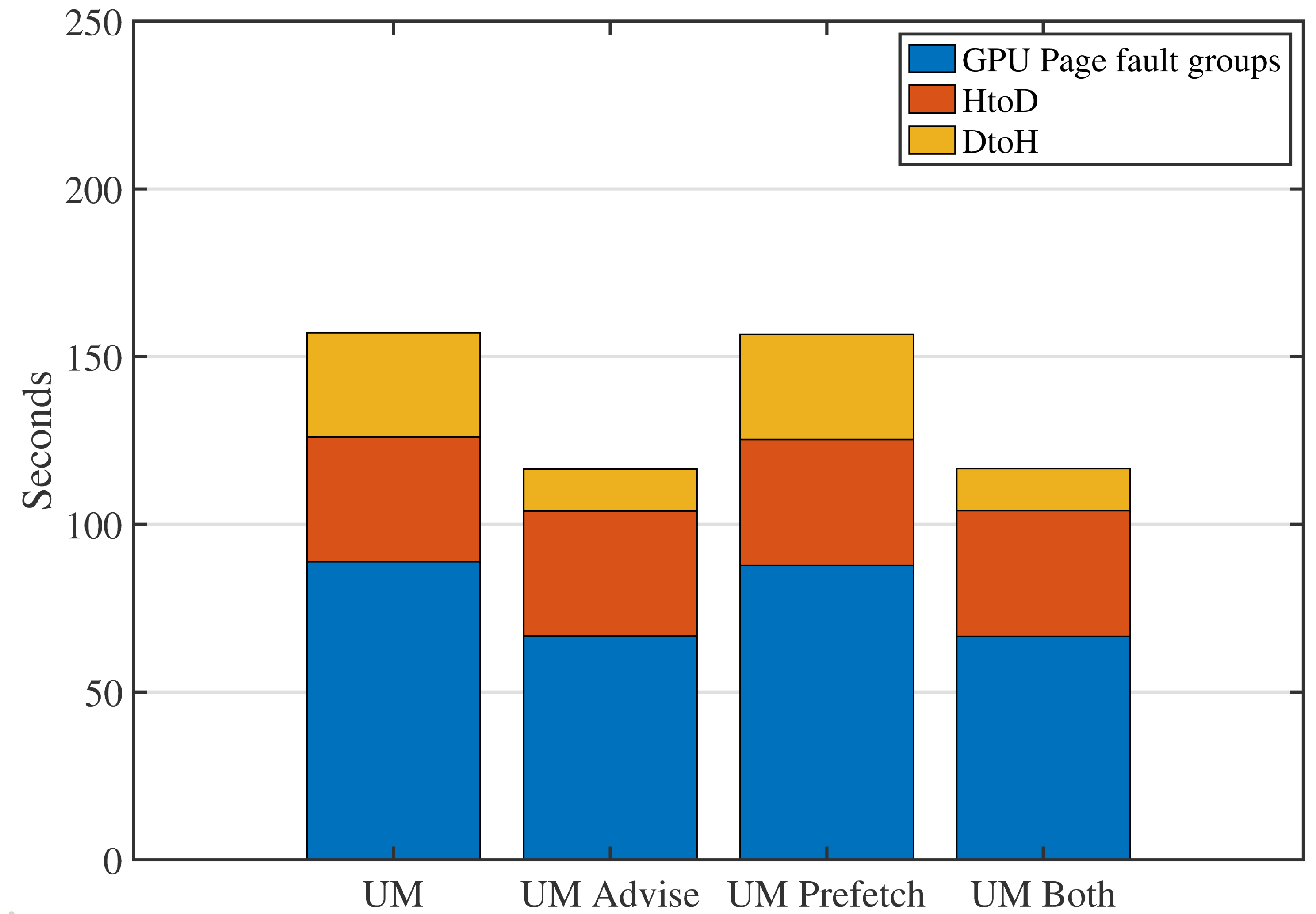}
		\caption{BS on Intel-Pascal}
		\label{fig:oversub-pagefault-breakdown-pascal-BS}
	\end{subfigure}
	~
	\begin{subfigure}[b]{0.2\textwidth}
		\includegraphics[width=\textwidth]{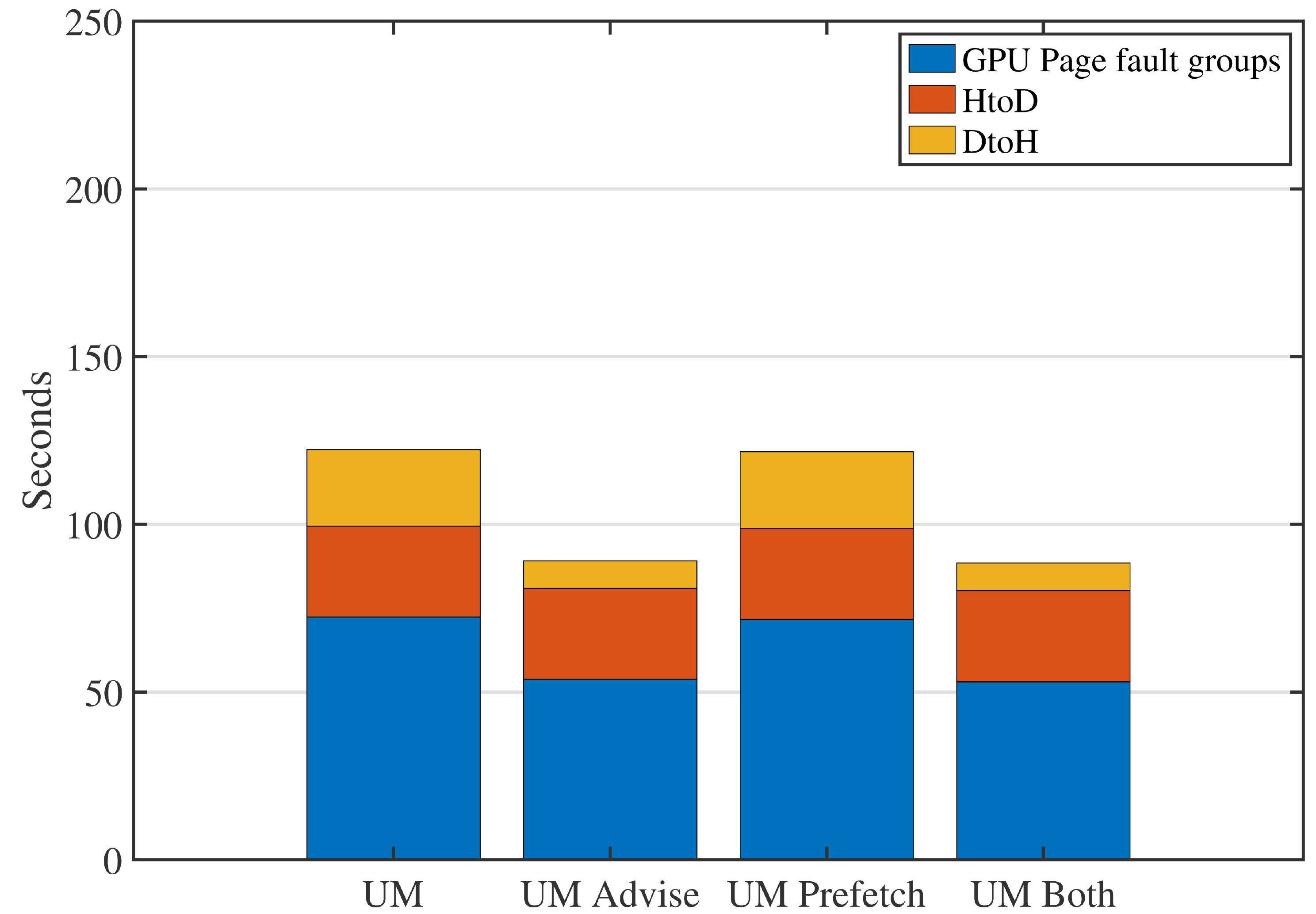}
		\caption{CG on Intel-Pascal}
		\label{fig:oversub-pagefault-breakdown-pascal-CG}
	\end{subfigure}
	~
	\begin{subfigure}[b]{0.2\textwidth}
		\includegraphics[width=\textwidth]{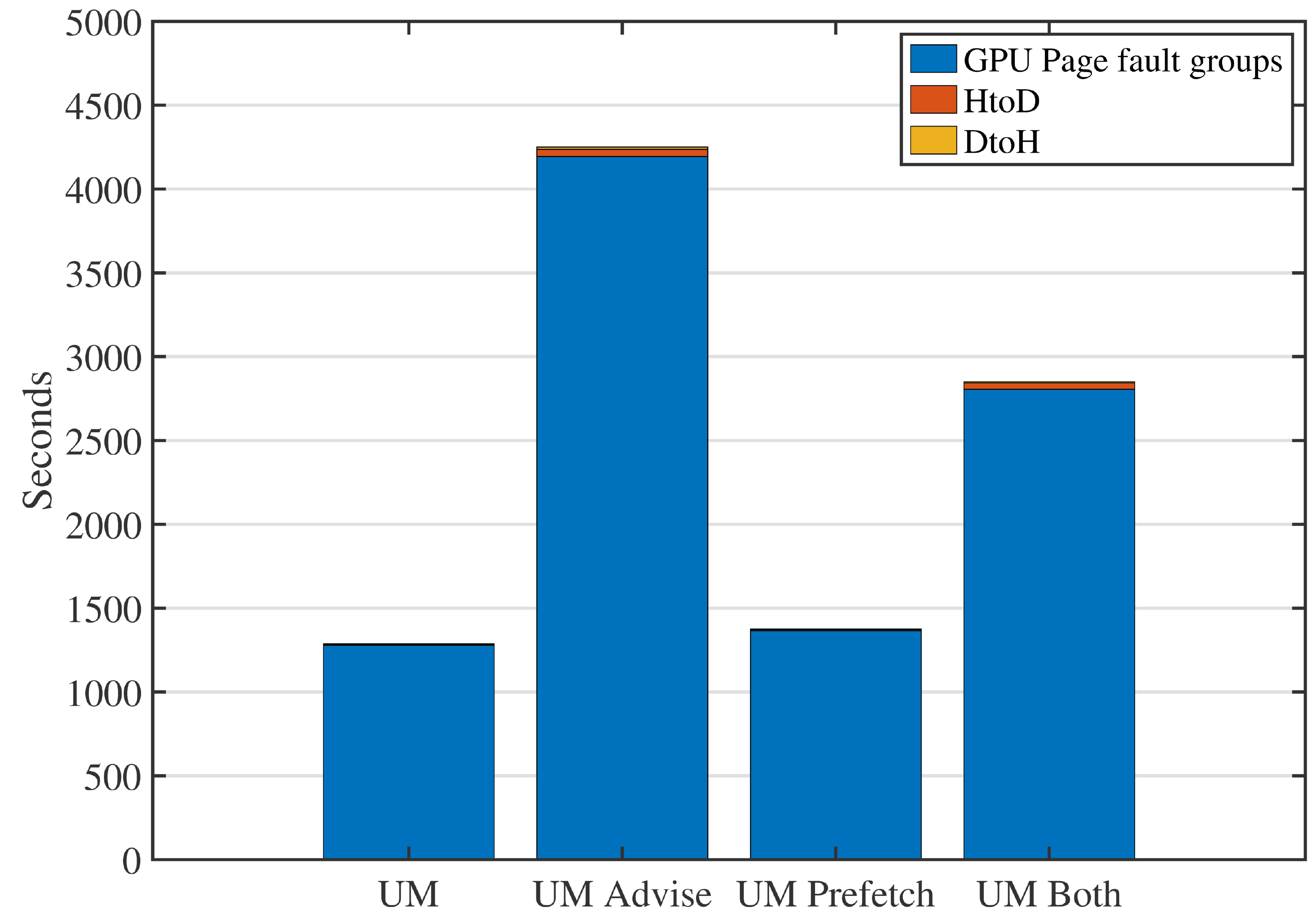}
		\caption{BS on P9-Volta}
		\label{fig:oversub-pagefault-breakdown-p9-BS}
	\end{subfigure}
	~
	\begin{subfigure}[b]{0.2\textwidth}
		\includegraphics[width=\textwidth]{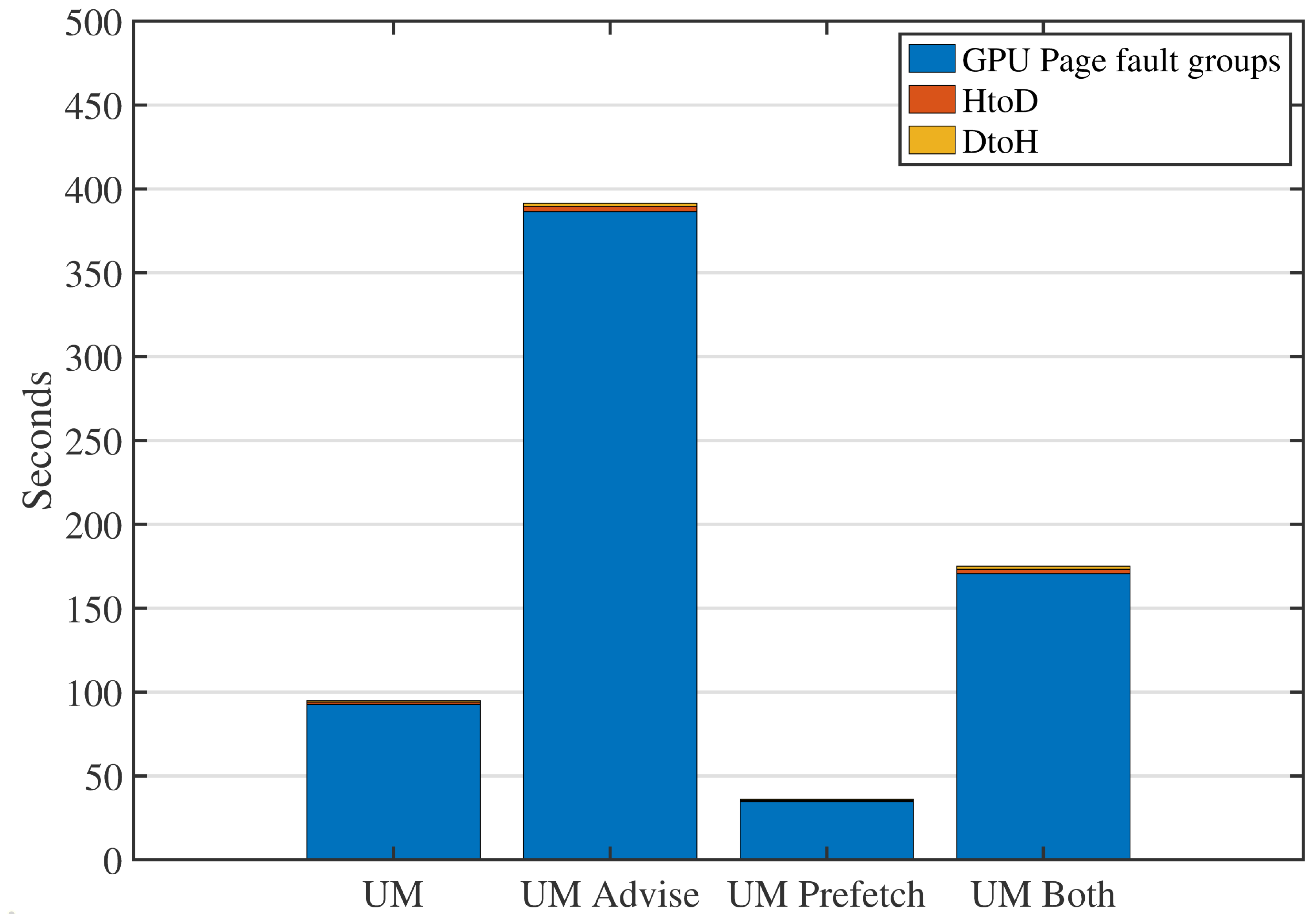}
		\caption{FDTD3d on P9-Volta}
		\label{fig:oversub-pagefault-breakdown-p9-FDTD3d}
	\end{subfigure}
	\caption{Breakdown of total time spent handling page faults and data movement when input size exceeds GPU memory.}
	\label{fig:pagefault-breakdown-oversubscribe}
\end{figure*}

\begin{figure*}
	\centering
	\begin{subfigure}[b]{0.225\textwidth}
		\includegraphics[width=\textwidth]{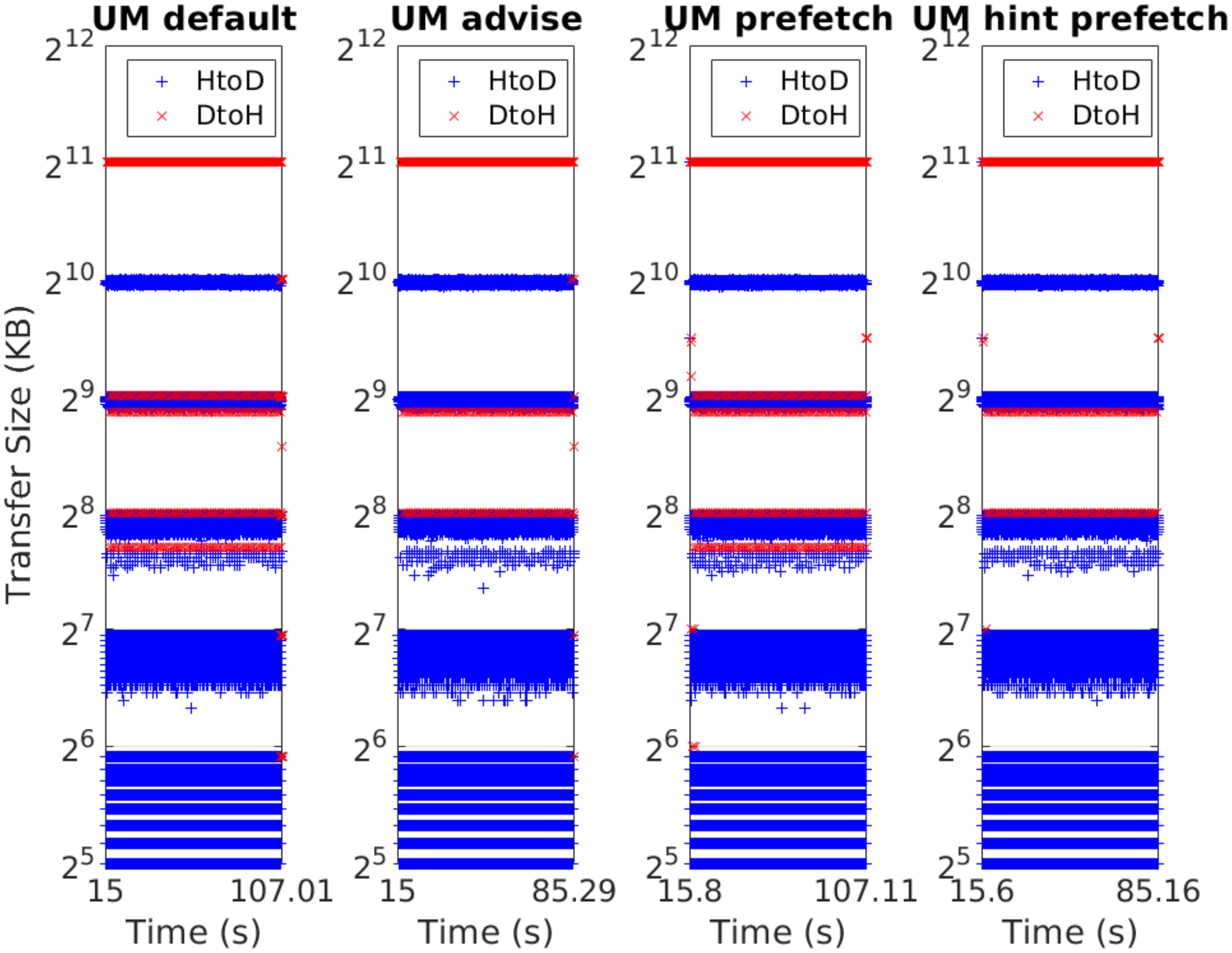}
		\caption{BS Intel-Pascal}
		\label{fig:trace-oversub-pascal-BS}
	\end{subfigure}
	~
	\begin{subfigure}[b]{0.225\textwidth}
		\includegraphics[width=\textwidth]{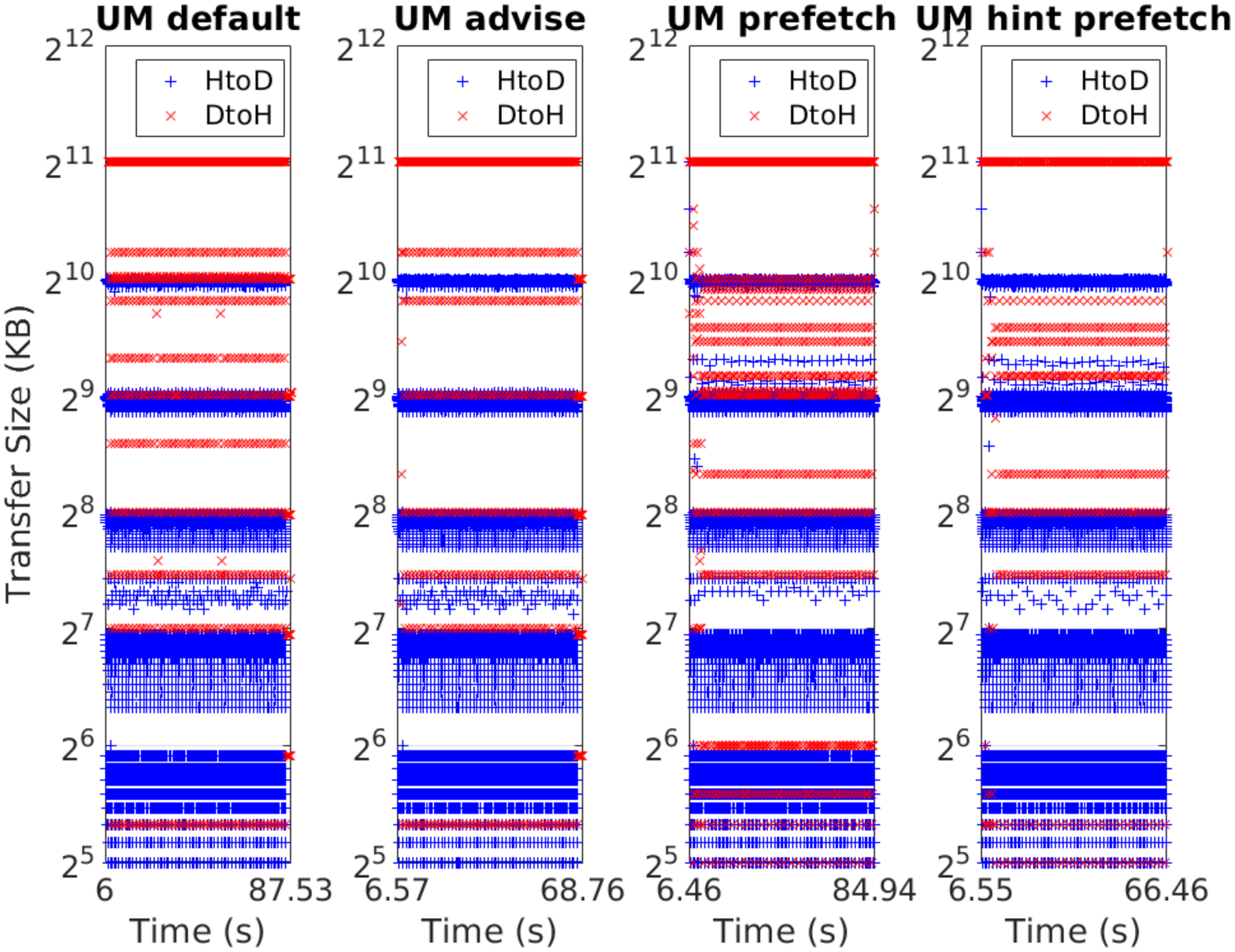}
		\caption{CG Intel-Pascal}
		\label{fig:trace-oversub-pascal-CG}
	\end{subfigure}
	~
	\begin{subfigure}[b]{0.225\textwidth}
		\includegraphics[width=\textwidth]{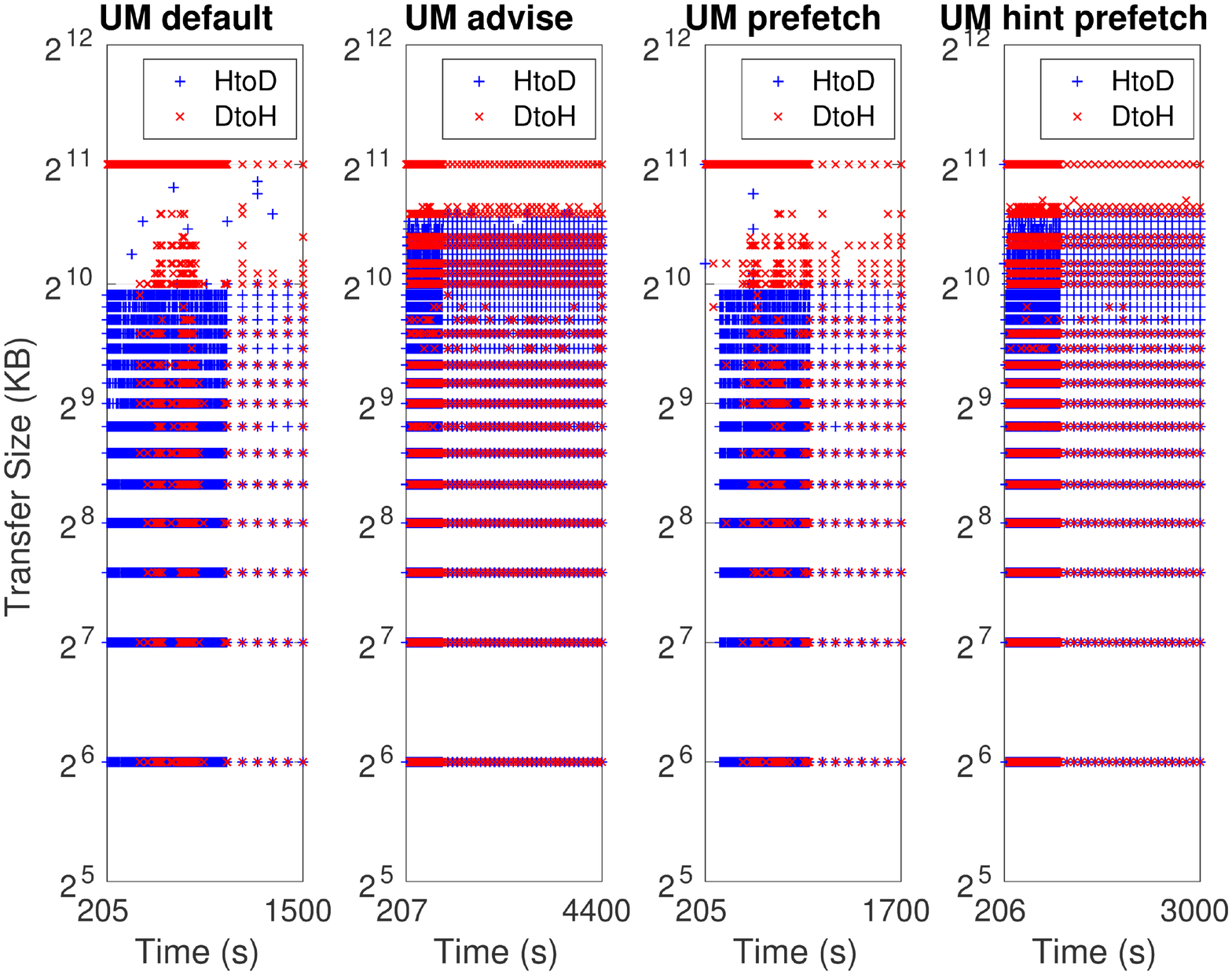}
		\caption{BS P9-Volta}
		\label{fig:trace-oversub-p9-BS}
	\end{subfigure}
	~
	\begin{subfigure}[b]{0.225\textwidth}
		\includegraphics[width=\textwidth]{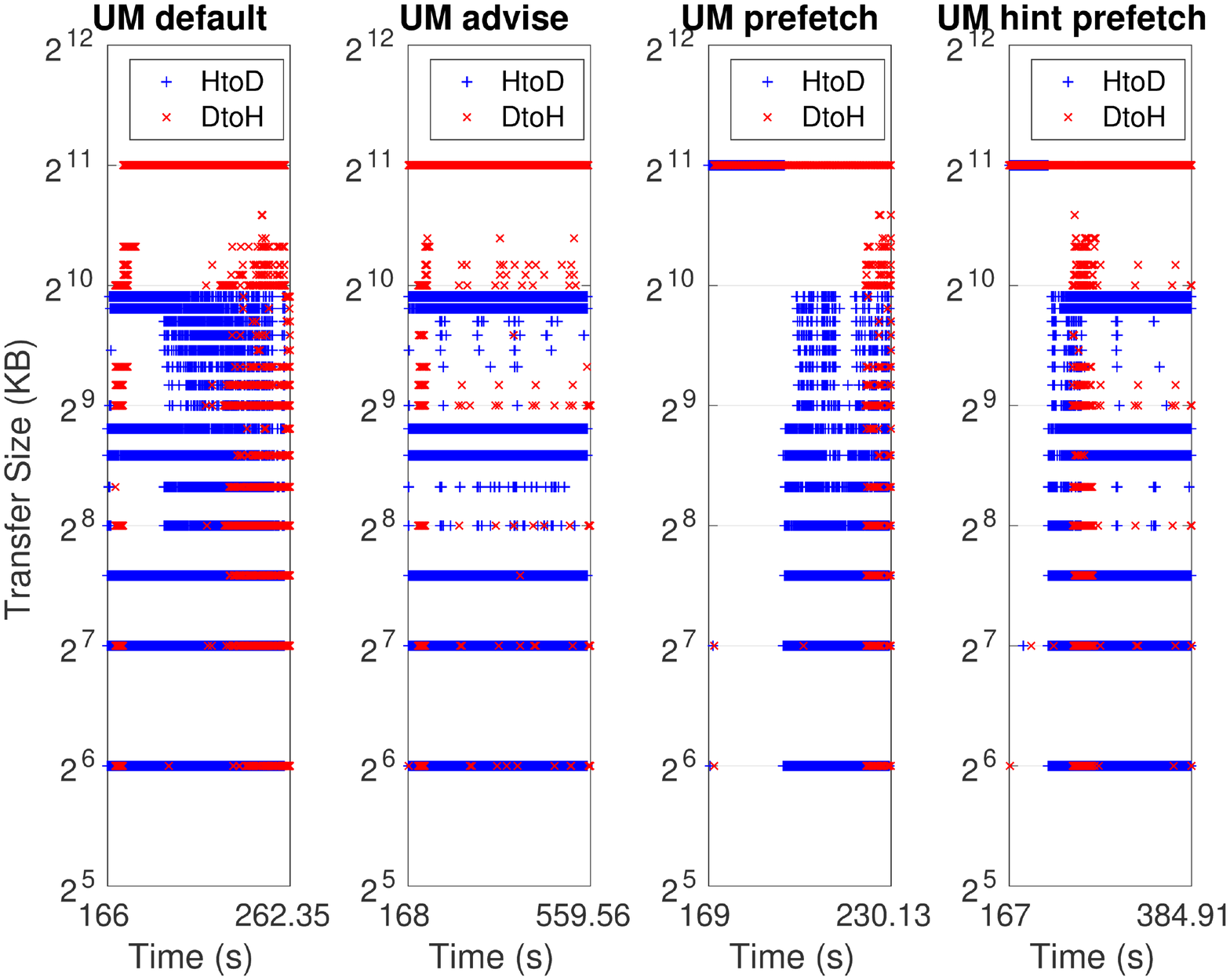}
		\caption{FDTD3d P9-Volta}
		\label{fig:trace-oversub-p9-FDTD3d}
	\end{subfigure}
	\caption{UM data transfer traces when input size exceeds GPU memory.}
	\label{fig:trace-oversubscribe}
\end{figure*}

Oversubscription of GPU memory is a key new feature of UM. It resembles the paging of unused memory pages to secondary storage to free up memory in classical virtual memory management. Similarly to the CPU memory subscription case, excessive use can lead to system slowdown and can severely impact performance. Our results show that all applications execute correctly, even when running out of GPU memory. However, techniques that improved performance for in-memory do not necessarily perform well when GPU memory is oversubscribed. On the contrary, the use of these techniques without careful optimization can lead to severe performance degradation.

We present the execution time of our applications in Fig.~\ref{fig:gpu-time-oversubscribe}. Since the case does not exist with original versions with explicit allocation, a comparison is not possible. Instead, the minimal UM version is used as a baseline. By using advise, specific applications can achieve up to over 20\% improvement on Intel platforms. Our P9 platform, on the other hand, shows a negative impact when advises are used. To better understand data movement, we perform tracing with the BS and CG on Intel-Pascal, and with BS and FDTD3d on P9-Volta.

For the Black-Scholes application, the use of advise results in performance improvement on Intel-Pascal. Fig.~\ref{fig:oversub-pagefault-breakdown-pascal-BS} shows the breakdown of time spent on page-fault related events of BS between host and device while Fig.~\ref{fig:trace-oversub-pascal-BS} shows the detailed tracing on Intel Pascal. One significant difference between default UM and UM advise is that a lot less time is spent on transferring data back to the host. The reduction in data movement can contribute to the improvement in performance. One possible reason for the reduction in data transfer from device to host is that instead of migrating data from GPU to host memory to make space, read-only data can simply be discarded as a copy already exists on host memory. On the other hand, on P9-Volta, significantly more time is spent on stalls. This can be seen in Fig.~\ref{fig:oversub-pagefault-breakdown-p9-BS}, where the total time is a few times higher than when no advise is used. Fig.~\ref{fig:trace-oversub-p9-BS} examines the data movement traces and clearly shows an intense data movement in both directions. This implies that the read-mostly advise has an interestingly negative effect on P9-Volta when data size exceeds device memory. A naive prefetch on Intel-Pascal provides performance improvement; however, it has little to no effect on P9-Volta.

CG on the Intel-Pascal platform benefits from using advise. The time breakdown for page faults and data movement is shown in Fig.~\ref{fig:oversub-pagefault-breakdown-pascal-CG}. As in the case of the Black-Scholes application, less time is spent on transferring data back to the host than in the case of basic UM. However, we note that a similar amount of data is sent from host to device in the two cases. This can also be seen in the detailed tracing in Fig.~\ref{fig:trace-oversub-pascal-CG}, where less device to host transfer is made.

FDTD3d is a finite difference solver, and it reads and writes to two arrays in an interleaving manner. Both arrays are being initialized using the same data. One of the arrays is being set to prefer GPU memory and will be accessed by the CPU. No advise is set on the other array. Since both arrays will be written to during execution, no read-mostly advise is set for them. However, read-mostly is set for a small array that contains coefficients. Fig.~\ref{fig:oversub-pagefault-breakdown-p9-FDTD3d} shows the breakdown of time in handling data movement and page faults on P9-Volta. Similarly to the Black-Scholes application, the usage of advise results in much higher spent on stalling. Execution time also increased significantly by approximately $3\times$. When prefetching, only one of those two data arrays is prefetched as they are originally identical. Interestingly, less data is seen transferring in both directions when prefetch is used. Fig.~\ref{fig:trace-oversub-p9-FDTD3d} shows the detailed tracing of the application. Smaller data transfers at the beginning become a bulk transfer. This is also reflected in the execution time, which is reduced from 60.9s to 45.3s as well as a reduction in time spent stalling. One possible reason is the size of the array being prefetched. Since only one array, which represents 50\% of the total problem size, is prefetched, the entire array can reside entirely on GPU memory without needed to evict previously prefetched data.
\section{Related Work}\label{sec:relwork}
The separate memory system between host and GPU has long been a programming challenge for developers. With UM, the runtime can transparently handle data movement between CPU and GPU. Earlier works~\cite{li2015evaluation, mishra2017benchmarking} have investigated the impact of UM in applications while ~\cite{mishra2017benchmarking} investigated the programming model support for UM in OpenMP through an extended LLVM compiler.  These studies lack the support of advanced memory features, which only become available recently. Recent efforts in the operating system, such as Heterogeneous Memory Management (HMM) in the Linux kernel~\cite{hmmRH, hmmLinux,sakharnykh2017unified}, provides mechanisms to mirror CPU page table on GPU and integrate device memory pages in the system page table by adding a new type of struct page.

CPU to GPU interconnect is another factor that impacts the performance of data movement directly. Extensive efforts have reported evaluation on modern GPU systems~\cite{li2018tartan, jia2018dissecting}. For instance, ~\cite{Pearson:2019:ECC:3297663.3310299}  developed a microbenchmark tool to evaluate the raw bandwidth performance with UM. While their works focus on interconnect performance and provide optimization insights, our work focuses on the impact of advanced memory features in optimizing the locality of pages.

Some of the recent works that apply advanced features of UM are Deep-Learning frameworks. One example is OC-DNN~\cite{awan2018oc}, an extended Caffe framework that uses UM to support the training of out-of-core batch sizes. They use memory advises to trigger data eviction and prefetch to trigger migration. They find these techniques useful in optimizing training performance. However, incorrect use can lead to performance degradation.

The memory oversubscription in GPU memory requires efficient page eviction to make space for newly requested pages. ~\cite{Ganguly:2019:IHP:3307650.3322224} proposed two pre-eviction policies using a tree-based neighborhood prefetching technique to select candidate pages.~\cite{Li2019Oversub} introduced an ETC framework for eager page pre-eviction and memory throttling in memory trashing. However, these optimization techniques target future GPU designs that require hardware modifications.
\section{Conclusion}
\label{sec:conclusion}

In this work, we investigated the impact of UM \emph{memory advises}, \emph{prefetch}, and GPU memory oversubscription, on CUDA application performance. We found that the performance of \emph{memory advises} mostly depends on the system in use and whether the GPU memory is oversubscribed. The use of \emph{memory advises} results in a performance improvement only when the GPU memory is oversubscribed on the Intel-Volta/Pascal-PCI-E systems. The use of \emph{memory advises} on Power9-Volta-NVLink based system, leads to a performance improvement when applications run in-memory while it results in a considerable performance degradation with GPU memory oversubscription. CUDA Unified \emph{prefetch} provides a performance improvement only on the Intel-Volta/Pascal-PCI-E based systems while it does not show a performance improvement on the Power9-Volta-NVLink based system.

In this work, we have set \emph{memory advises} for each memory object following best-practice guidelines from Nvidia. However, a future study on how to select optimal advise placement would help programmers derive different combinations of advises in different applications. In general, we found both \emph{memory advises} and \emph{prefetch} to be simple and effective. Overall, we showed that UM is a promising technology that can be used effectively when programming applications for GPU systems.

\section*{Acknowledgment}
\scriptsize{
	Funding for the work is received from the European Commission H2020 program, Grant Agreement No. 801039 (EPiGRAM-HS). Experiments were performed on resources provided by the Swedish National Infrastructure for Computing (SNIC) at HPC2N and Lassen supercomputer at LLNL. Part of this work was performed under the auspices of the U.S. Department of Energy by Lawrence Livermore National Laboratory under Contract DE-AC52-07NA27344 LLNL-PROC-788778. This research was also supported by the Exascale Computing Project (17-SC-20-SC), a collaborative effort of the U.S. Department of Energy Office of Science and the National Nuclear Security Administration.}

\bibliographystyle{plain}
\bibliography{main}

\end{document}